\documentclass{jfm}
\usepackage{graphicx}
\usepackage{epstopdf, epsfig}
\usepackage{bm}

\usepackage{amsmath,amssymb}
\usepackage{xcolor}
\usepackage{datetime}
\usepackage{subcaption}
\usepackage{setspace}
\usepackage[ruled,vlined,english]{algorithm2e}
\usepackage[babel=true]{csquotes}
\usepackage{mathrsfs}
\usepackage{amsbsy}

\shorttitle{MGEnKF-BARC}
\shortauthor{G. Moldovan et al.}

\title{Multigrid sequential data assimilation for the large-eddy simulation of a massively separated bluff-body flow}

\author{G. Moldovan\aff{1}
  \corresp{\email{gabriel.moldovan@ensma.fr}},
  A. Mariotti\aff{2},
  L. Cordier\aff{1},
  G. Lehnasch\aff{1},
  M. - V. Salvetti\aff{2}
 \and M.  Meldi\aff{3}}

\affiliation{\aff{1}Institut Pprime, CNRS, ENSMA, Université de Poitiers,
Département Fluides Thermique et Combustion, 86360 Futuroscope-Chasseneuil, France
\aff{2}DICI, Università di Pisa, Italy
\aff{3}Univ. Lille, CNRS, ONERA, Arts et Métiers ParisTech, Centrale Lille, UMR 9014- LMFL- Laboratoire de Mécanique des fluides de Lille - Kampé de Feriet, F-59000 Lille, France}

\begin{document}
\maketitle

\begin{abstract}
The potential for data-driven applications to scale-resolving simulations of turbulent flows is assessed herein. Multigrid sequential data assimilation algorithms have been used to calibrate solvers for Large Eddy Simulation for the analysis of the high-Reynolds-number flow around a rectangular cylinder of aspect ratio 5:1. This test case has been chosen because of a number of physical complexities which elude accurate representation using reduced-order numerical simulation.

The results for the statistical moments of the velocity and pressure flow field show that the data-driven techniques employed, which are based on the Ensemble Kalman Filter, are able to significantly improve the predictive features of the solver for reduced grid resolution. In addition, it was observed that, despite the sparse and asymmetric distribution of observation in the data-driven process, the data augmented results exhibit perfectly symmetric statistics and a significantly improved accuracy also far from the sensor location. 
\end{abstract}


\section{Introduction}

Large Eddy Simulation (LES) is currently one of the most used methods for the numerical prediction of turbulent flows \citep{Pope2000_cambridge,Sagaut2006_springer}. LES is based on the filtering of the flow scales and consists in simulating explicitly the large scales of the fluid motion and modelling the influence of the smallest scales. LES is available in both licensed and open-source codes.  It can intrinsically represent unstationary and three-dimensional flows via a time-and-space resolved approach, but with fewer computational resources when compared with Direct Numerical Simulation. Moreover, it can be easily coupled with RANS methods or wall-function to alleviate even further the computational needs for near-wall turbulence. These favourable features have extensively promoted its diffusion in the academic community and, more recently, for industrial applications.

However, the main limitation of this technique is that the theoretical framework for which the approach is developed can be rigorously reproduced only for academic test cases, i.e. simple geometries and uniform grids. In more complex cases, approximations can of course be implemented, but it may lead to unexpected results, such as lower accuracy of the results with mesh refinement. Numerous works in the literature \citep{Meyers2006_jfm,Meldi2012_pof,Ouvrard2010,Mariotti2017, Geurts2009, Meyers2003} indicate that these unexpected observations are tied with non-linear interactions between different sources of error, such as explicit and implicit filtering, discretization procedures, quality of the mesh, general set-up of the test case and subgrid scale modelling. These elements are usually governed by numerous free parameters that must be chosen by the user. Extensive analyses using classical trial and error approaches are in most cases not feasible due to the large costs of each LES. 

Therefore, there exist a number of works in the literature targeting a probabilistic description of the multidimensional parametric space of LES, in order to obtain information about the interactions between error sources. These works, which were mainly based on Uncertainty Quantification (UQ) and Uncertainty Propagation (UP), provided insights about the behaviour of LES in several applications such as isotropic turbulence \citep{Lucor2007_jfm,Meldi2011_pof}, mixing layers \citep{Meldi2012_pof}, turbulent channel \citep{Safta2017376,Rezaeiravesh2018,Rezaeiravesh2021} and bluff-body flows \citep{Mariotti2017,Rocchio2020}. 

More recently, Data Assimilation (DA) techniques \citep{Daley1991_cambridge} have been proposed for the analysis and optimization of LES. Among the works in the literature, one can find applications of variational tools \citep{Chandramouli_Memin_Heitz_JCP_2020} as well as reduced-order sequential tools \citep{Meldi2017_jcp} based on the Kalman Filter \citep{Kalman1960_jbe}. Recent applications of the ensemble Kalman filter (EnKF) \citep{Evensen1994,Evensen2009_Springer,Asch2016_SIAM} to the analysis of initial conditions for LES of reacting flows have been reported in the literature \citep{Labahn2019_pci}. A comprehensive analysis of data-driven applications to LES subgrid scale modelling for the plane channel flow has been performed by \cite{Mons2021_prf}, using a hybrid variational - EnKF approach. The DA tools allow to infer and optimize the parametric description of the LES problem, both in terms of boundary and initial conditions as well as for the subgrid scale modelling. The works reported in the literature have shown the potential of DA to improve the predictive capabilities of LES solvers, but extensive analyses for complex flow configurations are needed to assess the potential of the data-driven tools.   

In the present work, a recent computational strategy based on the EnKF, the multigrid ensemble Kalman Filter (MGEnKF, \citealt{Moldovan2021_jcp}), is used to augment LES in a practical application of engineering interest, namely the high-Reynolds-number flow around a rectangular cylinder of aspect ratio 5:1, which is the object of an international benchmark (BARC, \citealt{Bruno2014}). Despite the simple body geometry, the flow dynamics are complex, including flow separation at the upstream edges, mean-flow reattachment on the cylinder side and vortex shedding in the wake. Thus, BARC is a paradigmatic example of the intricate flows around elongated bodies often encountered in civil engineering applications, as, e.g., high buildings or bridge sections. BARC, which was conceived as a double-blind benchmark, i.e., without a priori selected reference measurements, collected a number of numerical and experimental studies. As shown in the first review work \citep{Bruno2014}, the LES predictions of some flow features and quantities are affected by a significant dispersion and by a high sensitivity to modelling and numerical parameters, such as subgrid scale modelling \citep{Mariotti2017} and grid refinement \citep{Bruno2012,Mariotti2017}. More recently, \cite{Rocchio2020} showed that the introduction of a very small rounding of the upstream edges, such as those due to manufacturing tolerances, has a strong impact on LES predictions. 


For all these reasons, this kind of flow is well adapted and challenging for the appraisal of DA procedures. The MGEnKF algorithm is used herein to improve the predictive capabilities of LES carried out on a rather coarse grid, by integrating reference data from a high-fidelity LES, run on a highly refined grid. More precisely, the MGEnKF algorithm will calibrate a parameter which regulates the dissipation introduced by a modal filter that can be considered as a subgrid scale dissipation. To the knowledge of the Authors, the present work is among the most advanced DA applications to scale-resolving numerical simulation to date.

The manuscript is organized as follows. In Section \ref{sec:num}, the numerical and data-driven tools employed for the analysis are introduced. In Section \ref{sec:RefSimulation}, the numerical results obtained from the high-fidelity LES are analysed and compared with results in the literature. In Section \ref{sec:DA-LES}, the DA procedure coupling the MGEnKF and the LES solver is described. The results obtained are discussed, in terms of state estimation and of parametric optimization of the numerical set-up. Finally, in Section \ref{sec:conclusions} concluding remarks are drawn.

\section{Numerical and modelling ingredients}
\label{sec:num}

\subsection{Numerical model: Nek$5000$}
\label{sec:num1}

The numerical simulations are performed using Nek5000, an open-source code based on a high-order spectral element method \citep{fischer2008} tailored for incompressible flow simulation. Each spectral element is rectangular or a suitable coordinate mapping of a rectangle. The basis functions inside the elements are Legendre polynomials of order $N$ for velocity and $N-2$ for pressure in each direction. Based on the high-order splitting method developed in \cite{mady1990}, a third-order backward finite-difference scheme is used for time advancing. The viscous terms are treated implicitly while the convective terms are explicit, with a third-order forward extrapolation in time. 

In Nek5000, when dealing with high Reynolds numbers, the spectral element method can be stabilized by applying a filtering method in the modal space. Here, a low pass filter is employed as in \cite{Mariotti2017} and \cite{Rocchio2020}. At the end of each step of the time integration, the low-pass explicit filter is applied to the velocity field in the modal space. This filter is characterized by a quadratic transfer function, which acts from the mode $N-k_c$ up to the highest mode $N$. The transfer function is written as follows:
\begin{equation}
	\begin{cases}
		\sigma_k=1 & k<k_c \\
  \displaystyle
		\sigma_k=1- w \left(\frac{k-k_c}{N-k_c}\right)^2 & k_c\le k\le N.
	\end{cases}
	\label{nek_filter}
\end{equation}
Once the cut-off mode ($k_c$) is set, it can only be tuned through a weighting parameter $w$. Since the filter acts only at the highest resolved modes, it may be interpreted as a SGS dissipation \citep{nek_sgs_1,nek_sgs_2}. Clearly, the behaviour of the stabilizing filter is affected by the choice of the cut-off mode $k_c$. However, in order to reduce the number of operative parameters and to be consistent with the analogous study carried out in \cite{Mariotti2017} for perfectly sharp upstream edges, $k_c$ is constant in this work ($k_c=3$ herein). 

\subsection{Test case and numerical set-up}
\label{sec:test_case}

\begin{figure}
	\centering
	{\includegraphics[width=0.85\textwidth]{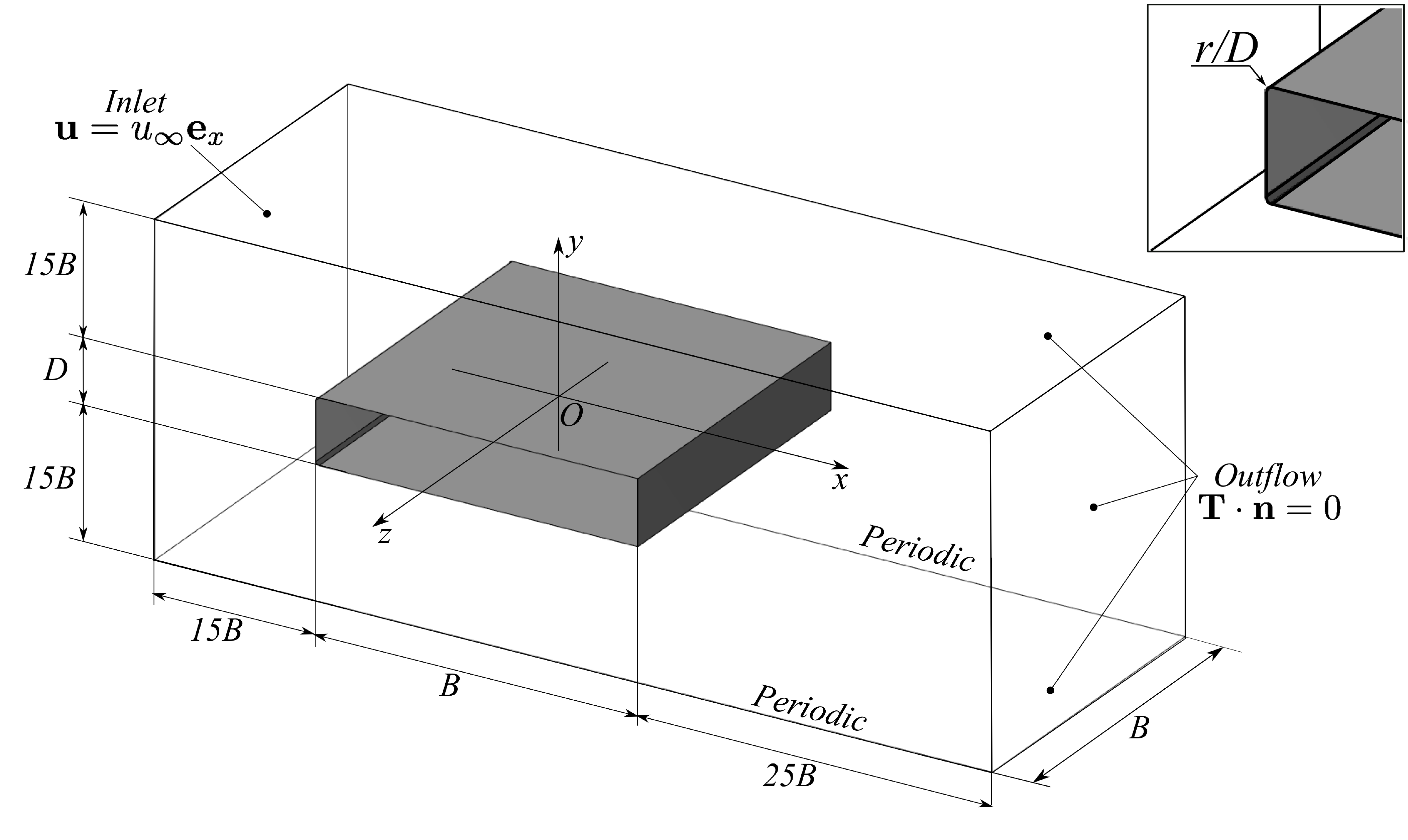}}\\
	\caption{Sketch of the computational domain.}
	\label{fig:domain}
\end{figure}

We consider the incompressible flow around an elongated rectangular cylinder, whose chord-to-depth ratio is $B/D=5$, at zero angle of attack. The cylinder upstream edges are rounded with a curvature radius $r/D=0.0037$, which is the smallest one considered in \cite{Rocchio2020}. This value falls within possible manufacturing tolerances. The computational domain is sketched in Figure \ref{fig:domain}, where $x$, $y$ and $z$ denote the streamwise, vertical and spanwise directions, respectively. The domain is rectangular and has the same dimensions as the ones used in other LES contributions to the benchmark \citep{Bruno2014,Bruno2012,Mariotti2017,Rocchio2020}. The centre of the cylinder is located at $x=y=0$. The computational domain is included in $-15.5\le x/B \le 25.5$, $-15.1\le y/B\le 15.1$ and the spanwise domain size is equal to B. A uniform velocity profile with no turbulence is imposed at the inlet. A no-slip velocity condition is adopted at the body surface, and traction-free boundary conditions are used for the outflow and for the upper and lower boundaries of the domain. Finally, periodicity is imposed in the spanwise direction. The Reynolds number, $Re$, based on the free-stream velocity, $u_\infty$, and on the cylinder depth, $D$, is $4\cdot 10^4$. 

Several numerical simulations of this test case, which will be discussed in the next sections, have been performed in this work using three different grids. Information about each of the meshes is now provided:
\begin{enumerate}
    \item the high-resolution \textit{refined grid}, for which the size of the elements in the spanwise direction is constant and its value is $\Delta z/D = 0.275$. The near wall resolution in the streamwise and vertical directions is $\Delta x/D=\Delta y/D=0.0625$. The meshing strategy is the same as the one used by \cite{Bruno2014,Bruno2012,Mariotti2017,Rocchio2020}.
    \item The \textit{baseline grid} shows a very similar structure to the \textit{refined grid}, but it is significantly coarser. The resolution in the spanwise direction is $\Delta z/D = 0.346$ and in the streamwise and vertical direction is $\Delta x/D=\Delta y/D=0.125$. The total number of mesh elements for this simulation is approximately $10^7$, i.e. around 10 times less than the number in the \textit{refined grid}.
    \item The \textit{coarse grid} is the last mesh used in this study, and it is the coarsest one. The resolution is $\Delta z/D = 1, \; \Delta x/D=\Delta y/D=0.25$ for a total of approximately $2 \times 10^6$ mesh elements.
\end{enumerate}
  The details about the mesh employed in this work are summarized in Table \ref{tab:setupMesh}. 
  In all cases, the same strategy as in \citep{Rocchio2020} is adopted to discretize the region close to the rounded upstream edges. In particular, the rounded part is handled by curvilinear elements and a transition box, which allows to move from a Cartesian grid to
a Cylindrical one, is used to transform each spectral element containing the upstream rounding edges.

Inside each spectral element, the order of Legendre polynomials is set to $N=6$, as in previous published works \citep{Mariotti2017,Rocchio2020,Lunghi2022}.

We stress again that, for every simulation, the cutoff threshold for the filter in the modal space described in Sec. \ref{sec:num1} is set to $K_c=3$ as in \cite{Mariotti2017,Rocchio2020}. The filter weight for the \textit{reference} and \textit{baseline} simulations is $w=0.018$. This value is lower than that adopted in the LES in \cite{Rocchio2020} and at the lowest bound of the values considered in \cite{Mariotti2017}. This choice is motivated by the fact that the variability of the results is expected to be larger for low values of $w$, and thus, of SGS dissipation, as confirmed by the results in \cite{Mariotti2017} and by preliminary simulations for rounded edges. On the other hand, for low values of $w$ and refined grids, \cite{Mariotti2017} showed that LES simulations give a solution which is in disagreement with experimental data. However, since we wish to assess herein the capabilities of a data driven approach to improve the accuracy of LES solutions on coarse grids, we prefer to enhance the variability of the solution with grid refinement by adopting a low value of $w$, even if this choice is not optimal for the agreement with the experimental data, and we will consider the solution on the refined grid as the \textit{ground truth}. 

	\begin{table}
		\centering
		\begin{tabular}{ccccc} \hline
			Grid	  &  $\Delta x/D$ & $\Delta y/D$ & $\Delta z/D$ & Nr elements \\ \hline
			Refined grid     & $0.0625$ & $0.0625$ & $0.275$ & $\approx10^8$\\
			Baseline grid & $0.125$ & $0.125$ & $0.558$ & $\approx10^7$ \\
			Coarse grid & $0.25$ & $0.25$ & $1$ & $\approx 10^6$ 
		\end{tabular}
		\caption{\label{tab:setupMesh} Details about the grid resolutions.}
	\end{table}

\begin{table}
	\centering
	\begin{tabular}{ccc} \hline
	        Simulation  & grid & $w$  \\ \hline
			\textit{Reference LES} & refined grid & $0.018$ \\
			\textit{Baseline LES} & baseline grid & $0.018$ \\
   			\textit{Data-augmented LES} & baseline grid ($1$ sim.) &  \\
      	 & + coarse grid ($N_e$ sim.) & $0.018 \to 0.00623$ \\
			\textit{Optimized LES} & baseline grid & $0.00623$
	\end{tabular}
	\caption{\label{tab:setupLES} Details about the numerical simulations performed in this work.}
\end{table}

\subsection{KF, EnKF and MGEnKF}

The Kalman Filter (KF), firstly introduced by \citep{Kalman1960_jbe}, is a sequential data assimilation tool which provides an estimation of a state variable $\mathbf{x}_k$ at time $k$. This state estimation is obtained by combining a prior state $\mathbf{x}_{k-1}$ at a time $t_{k-1}$, a set of observations $\mathbf{y}^\text{o}_k$ and a linear dynamical model $\mathbf{M}_{k:k-1}$. The estimation is obtained via two successive operations: \textit{forecast} (superscript $f$) and \textit{analysis} (superscript $a$). A first forecast of the state of the system $\mathbf{x}^\text{f}_{k}$ and the error covariance $\mathbf{P}^\text{f}_{k}$ is obtained as follows:

\begin{subequations}
\begin{align}
\mathbf{x}^\text{f}_{k} & = \mathbf{M}_{k:k-1} \mathbf{x}^\text{a}_{k-1}\label{eq:x_mdl_kf}\\
\mathbf{P}^\text{f}_{k}& = \mathbf{M}_{k:k-1} \mathbf{P}^\text{a}_{k-1} {\mathbf{M}^\top_{k:k-1}}+\mathbf{Q}_k. \label{eq:P_mdl_kf}
\end{align}
\end{subequations}

The state and covariance are then updated with the available observation $\mathbf{y}^\text{o}_k$ via the Kalman Gain, $\mathbf{K}_k$, a weighting matrix that reflects the error present in the model ($\mathbf{Q}_k$) and the error present in observations ($\mathbf{R}_k$). The required matrix operations for the analysis phase can be written as follows:

\begin{subequations}
\begin{align}
  \mathbf{K}_k& = \mathbf{P}^\text{f}_{k}{\mathbf{H}^\top_k}\left(\mathbf{H}_k \mathbf{P}^\text{f}_{k} {\mathbf{H}^\top_k}+\mathbf{R}_k\right)^{-1}\label{eq:k_kf}\\
  \mathbf{x}^\text{a}_{k}& = \mathbf{x}^\text{f}_{k} + \mathbf{K}_k\left(\mathbf{y}^\text{o}_k-\mathbf{H}_k\mathbf{x}^\text{f}_{k}\right)\label{eq:x_kf}\\
  \mathbf{P}^\text{a}_{k}& = \left(I-\mathbf{K}_k \mathbf{H}_k\right)\mathbf{P}^\text{f}_{k}\label{eq:P_kf}
\end{align}
\end{subequations}


The matrix operations in \eqref{eq:P_mdl_kf}, \eqref{eq:P_kf} and in particular \eqref{eq:k_kf} are extremely expensive for realistic CFD applications, and they may require computational resources that are orders of magnitude larger than the model operation. In order to alleviate such computational issues, one possible alternative is to provide an approximation of $\mathbf{P}$ using stochastic Monte-Carlo sampling. The Ensemble Kalman Filter (EnKF) \citep{Evensen1994,Evensen2009_Springer} provides an approximation of $\mathbf{P}_k$ by means of an ensemble of $N_\text{e}$ states.

Given an ensemble of forecast/analysis states at a given instant $k$, the ensemble matrix collecting the information of the database is defined as:
\begin{equation}
\pmb{\mathscr{E}}_k^{\text{f}/\text{a}}=
\left[\mathbf{x}_k^{\text{f}/\text{a},(1)},\cdots,\mathbf{x}_k^{\text{f}/\text{a},(N_\text{e})}\right]\in\mathbb{R}^{N_x\times N_\text{e}}, \label{eq:Ensemble_Matrix}
\end{equation}

where $N_x$ is the size of the state vector. Starting from the data assembled in $\pmb{\mathscr{E}}_k^{\text{f}/\text{a}}$, the ensemble mean 
\begin{equation}
    \overline{\mathbf{x}_k^{\text{f}/\text{a}}}=\frac{1}{N_\text{e}}\sum^{N_\text{e}}_{i=1}\mathbf{x}_k^{\text{f}/\text{a},(i)} \label{eq:Ensemble_Mean}
\end{equation} 

is used to obtain the normalized ensemble anomaly matrix:
\begin{equation}
\mathbf{X}_k^{\text{f}/\text{a}}=\frac{\left[\mathbf{x}_k^{\text{f}/\text{a},(1)}-\overline{\mathbf{x}_k^{\text{f}/\text{a}}},\cdots,\mathbf{x}_k^{\text{f}/\text{a},(N_\text{e})}-\overline{\mathbf{x}_k^{\text{f}/\text{a}}}\right]}{\sqrt{N_\text{e}-1}}\in\mathbb{R}^{N_x\times N_\text{e}}, \label{eq:Ensemble_Anomaly}
\end{equation}

The approximated error covariance matrix, hereafter denoted with the superscript $e$, may be obtained via the matrix product: 
\begin{equation}
\mathbf{P}_k^{\text{f}/\text{a},\text{e}}=\mathbf{X}_k^{\text{f}/\text{a}} \left(\mathbf{X}_k^{\text{f}/\text{a}}\right)^\top\in\mathbb{R}^{N_x\times N_x} \label{eq:Ensemble_P},
\end{equation}

although in practice, this calculation can be avoided. Concerning the observation data, \cite{Burgers1998} showed that to mimic the Best Linear Unbiased Estimator (BLUE) analysis of the Kalman Filter, it is necessary to consider $\mathbf{y}^\text{o}_k\in \mathbb{R}^{N_y}$ as a random variable following the statistics of the observation. It is thus necessary to define an ensemble of perturbed observations as:

\begin{equation}
\mathbf{y}_k^{\text{o},(i)}=\mathbf{y}_k^\text{o}+\mathbf{\epsilon}_k^{\text{o},(i)},
\quad
\text{with}
\quad
i=1,\cdots,N_\text{e}
\quad
\text{and}
\quad
\mathbf{\epsilon}_k^{\text{o},(i)}\sim \mathcal{N}(0,\mathbf{R}_k)
.\label{eq:perturbed_y}
\end{equation}

One can define the normalized anomaly matrix of the observations errors as
\begin{equation}
\mathbf{E}_k^{\text{o}}=
\frac{1}{\sqrt{N_\text{e}-1}}
\left[
\mathbf{\epsilon}_k^{\text{o},(1)}-\overline{\mathbf{\epsilon}_{k}^{\text{o}}},
\mathbf{\epsilon}_k^{\text{o},(2)}-\overline{\mathbf{\epsilon}_{k}^{\text{o}}},
\cdots,
\mathbf{\epsilon}_k^{\text{o},(N_\text{e})}-\overline{\mathbf{\epsilon}_{k}^{\text{o}}},
\right]
\in\mathbb{R}^{N_y\times N_\text{e}},
\end{equation} 
where
$\displaystyle\overline{\mathbf{\epsilon}_{k}^{\text{o}}}=\frac{1}{N_\text{e}}\sum_{i=1}^{N_\text{e}}\mathbf{\epsilon}_k^{\text{o},(i)}$.

The covariance matrix of the measurement error can then be estimated as
\begin{equation}
\mathbf{R}_k^\text{e}=\mathbf{E}^\text{o}_k \left(\mathbf{E}^\text{o}_k\right)^\top\in\mathbb{R}^{N_y\times N_y}. \label{eq:observation_matrix_errorcov}
\end{equation}


The analysis step is performed by updating each member of the ensemble as follows:
\begin{equation}
\mathbf{x}_k^{\text{a},(i)}=
\mathbf{x}_k^{\text{f},(i)}+
\mathbf{K}_k^\text{e}
\left(y_k^{\text{o},(i)}-\mathbf{\mathcal{H}}_k\left(\mathbf{x}^{\text{f},(i)}_k\right)\right)\label{eq:ensemble_update},
\end{equation}
where $\mathbf{K}_k^\text{e}$ is the Kalman Gain and $\mathbf{\mathcal{H}}_k$ is the non-linear observation operator. The Kalman Gain is approximated using ensemble statistics as
\begin{equation}
\mathbf{K}_k^\text{e}=
\mathbf{X}_k^\text{f}
\left(\mathbf{Y}_k^\text{f}\right)^\top
\left(
\mathbf{Y}_k^\text{f}
\left(\mathbf{Y}_k^\text{f}\right)^\top
+
\mathbf{E}_k^\text{o} \left(\mathbf{E}_k^\text{o}\right)^\top
\right)^{-1},
\end{equation}
where $\mathbf{Y}_k^\text{f}=\mathbf{H}_k\mathbf{X}_k^\text{f}$.



The EnKF has been used in several fluid mechanics and engineering applications, such as wildfire propagation \citep{Rochoux2014_nhess}, combustion \citep{Labahn2019_pci}, turbulence modelling \citep{Xiao2016_jcp,Zhang2020_cf,Zhang2021_cf} and hybrid variational-EnKF methods \citep{Mons2019_jcp}. Parameter estimation procedures can be easily implemented to the EnKF framework. Several alternatives exist (see \citealt{Asch2016_SIAM}), amongst which we have selected the \textit{dual estimation} strategy proposed by \cite{moradkhani2005}.


\subsection{Multigrid Ensemble Kalman Filter (MGEnKF)}
The size of the ensemble required to correctly represent the uncertainty in the model prediction can easily be a limiting factor. The computational expenses required to have ensemble sizes between $40-100$ members for complex fluid mechanics  estimation problems result in computationally intractable applications. 
This is the reason several research works on reduced-order variations of classical DA methods have been recently proposed in the literature. Among those, the Multigrid Ensemble Kalman Filter (MGEnKF) \citep{Moldovan2021_jcp, Moldovan2022} combines a single simulation performed on a fine grid with an ensemble simulation run on coarser grid levels. The statistics obtained on the coarse grid are used to update the fine grid simulation, which is in turn used to reduce the numerical errors present in the ensemble. Fine and coarse grid simulations are carried out in a synergistic way, allowing for computationally tractable applications of the EnKF to complex fluid mechanics problems. The proposed algorithm is inspired from Multigrid resolution methods, and more precisely, the Full Approximation Scheme method described in \cite{Brandt1977_mc,Wesseling1999_jcam}.

\begin{figure}[htbp]
\centering
\includegraphics[width=0.95\textwidth]{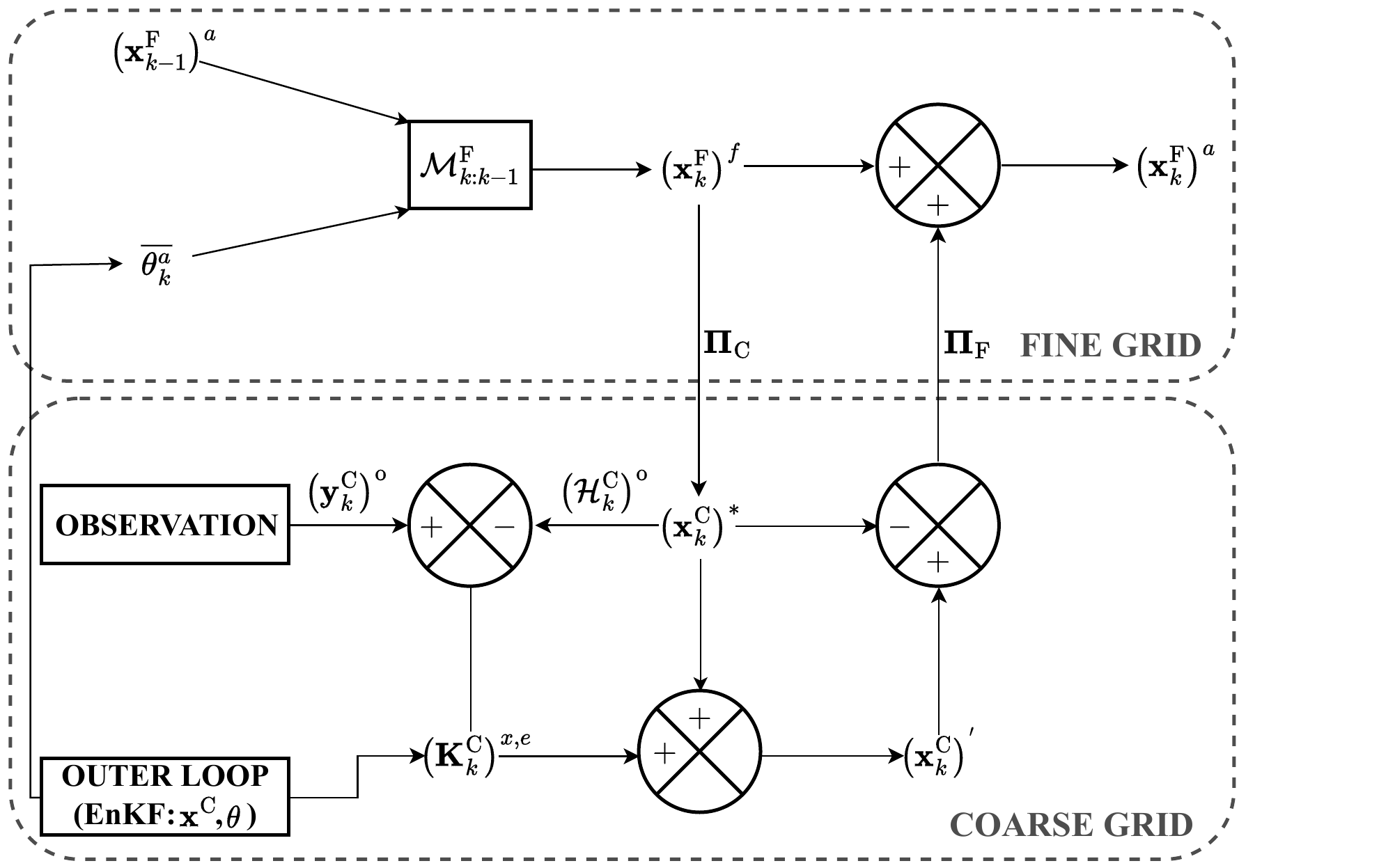}
\caption{\label{fig:schema_MGENKF}
Schematic representation of the Multigrid Ensemble Kalman Filter (MGEnKF) by \cite{Moldovan2021_jcp}. Two different levels of representation (fine and coarse grids) are used to obtain an estimation for the main simulation running on the fine grid.}
\end{figure}

\begin{figure}
\centering
\includegraphics[width=1\textwidth]{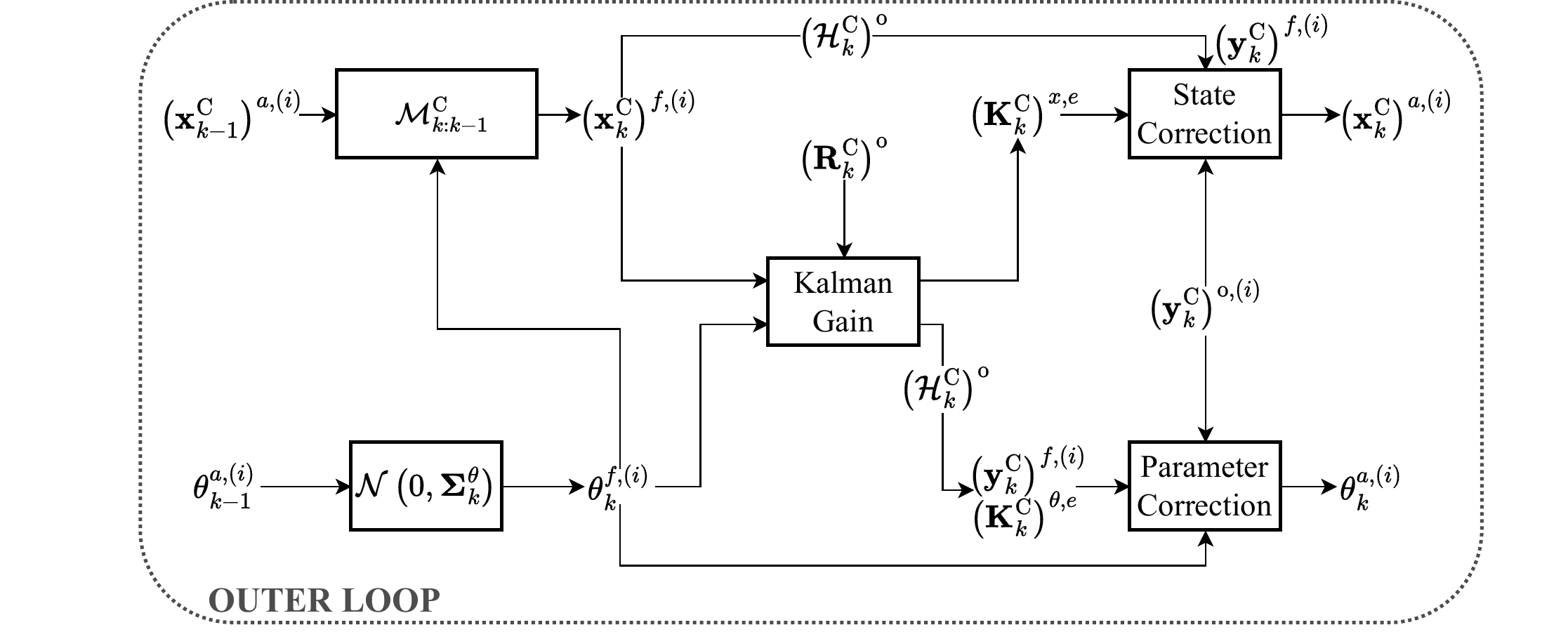}
\caption{\label{fig:schema_inou_loop}
Schematic representation of the \textit{outer loop} box in Fig.~\ref{fig:schema_MGENKF}.
}
\end{figure}

The scheme of a two-grid application of the MGEnKF is shown in Fig.~\ref{fig:schema_MGENKF}. Herein, the version of the model proposed in the work by \cite{Moldovan2021_jcp} is used.
This method reduces the computational costs of the data assimilation procedure and provides solutions which are compatible with the dynamic equations of the model. 
The MGEnKF algorithm is structured in the following operations:

\begin{enumerate}
\item \textbf{Predictor step}. A forecast state $\left(\mathbf{x}^\text{\tiny F}_{k}\right)^\text{f}$ is obtained from the initial solution on the fine grid $\left(\mathbf{x}^\text{\tiny F}_{k-1}\right)^\text{a}$  
$$
	\left(\mathbf{x}_k^\text{\tiny F}\right)^\text{f}=
	\mathbf{\mathcal{M}}^\text{\tiny F}_{k:k-1}\left(
	\left(\mathbf{x}_{k-1}^\text{\tiny F}\right)^\text{a},
	\overline{\mathbf{\theta}_{k}^\text{a}}\right),
$$ 
where $\mathbf{\mathcal{M}}^\text{\tiny F}_{k:k-1}$ is the model parametrized by  $\displaystyle \overline{\mathbf{\theta}_{k}^\text{a}} = \frac{1}{N_\text{e}}\sum_{i=1}^{N_\text{e}}\mathbf{\theta}_{k}^\text{a,(i)}$, the ensemble average of coarse-level parameters.
Each member $i$ of the ensemble defined on the coarse grid is also advanced in time
$$
	\left(\mathbf{x}_k^\text{\tiny C}\right)^\text{f,(i)}=
	\mathbf{\mathcal{M}}^\text{\tiny C}_{k:k-1}\left(
	\left(\mathbf{x}_{k-1}^\text{\tiny C}\right)^\text{a,(i)},
	\mathbf{\theta}_{k}^\text{f,(i)}\right)
$$ 
where $\mathbf{\mathcal{M}}^\text{\tiny C}_{k:k-1}$ is the same model but used on the coarse grid.

%
\item \textbf{Projection on the coarse grid}. $\left(\mathbf{x}^\text{\tiny C}_{k}\right)^{*}$ is obtained by projecting $\left(\mathbf{x}^\text{\tiny F}_{k}\right)^\text{f}$ on the coarse grid with a projection operator $\Pi_\text{\tiny C}$, \textit{i.e.} 
$$
\left(\mathbf{x}^\text{\tiny C}_k\right)^{*}=
\Pi_\text{\tiny C}\left(\left(\mathbf{x}_k^\text{\tiny F}\right)^\text{f}\right).
$$
%
%
\item \textbf{\textit{Outer} loop (Data Assimilation)}.
If external observation $\left(\mathbf{y}_k^\text{\tiny C}\right)^\text{o}$ is available, the ensemble forecast $\left(\mathbf{x}^\text{\tiny C}_{k}\right)^{\text{f},(i)}$ and the model parameters $\mathbf{\theta}_{k}^{\text{f},(i)}$ are updated with the available observation $\left(\mathbf{y}_k^\text{\tiny C}\right)^\text{o}$ to obtain $\left(\mathbf{x}^\text{\tiny C}_{k}\right)^{\text{a},(i)}$ and $\mathbf{\theta}_{k}^{\text{a},(i)}$. 
\item \textbf{Determination of the state variables on the coarse grid}. The fine grid state of the system is updated on the coarse grid exploting multigrid features. This coarse grid update, referred to as $\left(\mathbf{x}^\text{\tiny C}_k\right)^{'}$, is obtained using the Kalman gain matrix $\left(\mathbf{K}_k^\text{\tiny C}\right)^{x,\text{e}}$ calculated in step \textit{(iii)} to perform a classical KF correction step, \textit{i.e.} 
		\begin{align*}
\left(\mathbf{x}^\text{\tiny C}_k\right)^{'} & =
\left(\mathbf{x}^\text{\tiny C}_k\right)^{*}+
\left(\mathbf{K}_k^\text{\tiny C}\right)^{x,\text{e}}
\left[
\left(\mathbf{y}_k^\text{\tiny C}\right)^\text{o}-
\left(\mathbf{\mathcal{H}}_k^\text{\tiny C}\right)^\text{o}
\left(\left(\mathbf{x}_k^\text{\tiny C}\right)^{*}\right)
\right]
		\end{align*} 
If observation is not available, the solution on the coarse grid is obtained using classical iterative procedures on the coarse grid, if a  multigrid solver is used.
%

\item \textbf{Final iteration on the fine grid}. The fine grid state solution $\left(\mathbf{x}^\text{\tiny F}_k\right)^{'}$ is determined using the results obtained on the coarse space: $\left(\mathbf{x}^\text{\tiny F}_k\right)^{'}=\left(\mathbf{x}^\text{\tiny F}_{k}\right)^\text{f}+\Pi_\text{\tiny F}\left(\left(\mathbf{x}^\text{\tiny C}_k\right)^{'}-\left(\mathbf{x}^\text{\tiny C}_{k}\right)^{*}\right)$. The state $\left(\mathbf{x}^\text{\tiny F}_k\right)^\text{a}$ is obtained from a final iterative procedure starting from $\left(\mathbf{x}^\text{\tiny F}_k\right)^{'}$. 
\end{enumerate}

A detailed representation of the \textit{Outer} loop box shown in Fig.~\ref{fig:schema_MGENKF} is provided in Fig.~\ref{fig:schema_inou_loop}. This method is reminiscent of multilevel \citep{Hoel2016_SIAM,Siripatana2019_cg,KodyLaw2020,Fossum2020_cg} and multifidelity \citep{Gorodetsky2020_jcp,Popov2021_SIAM} ensemble techniques for data assimilation. However, only one main simulation on the fine grid is needed, significantly reducing the computational cost associated to the whole procedure.


\section{Instantaneous and statistical features obtained by the \textit{reference LES}}
\label{sec:RefSimulation}

The results obtained with the \textit{reference LES} are now discussed and compared with the ones for the \textit{baseline LES} and with the ensemble statistics of the experimental and numerical contributions to the BARC benchmark reviewed in \cite{Bruno2014}. The \textit{reference LES} is the only simulation performed on the \textit{refined grid} (see Table \ref{tab:setupLES} for details). Its results are used to validate the other LES performed, and to serve as observation in the data-driven procedure. 

\begin{figure}
		\centering
		\includegraphics[width=.48\textwidth]{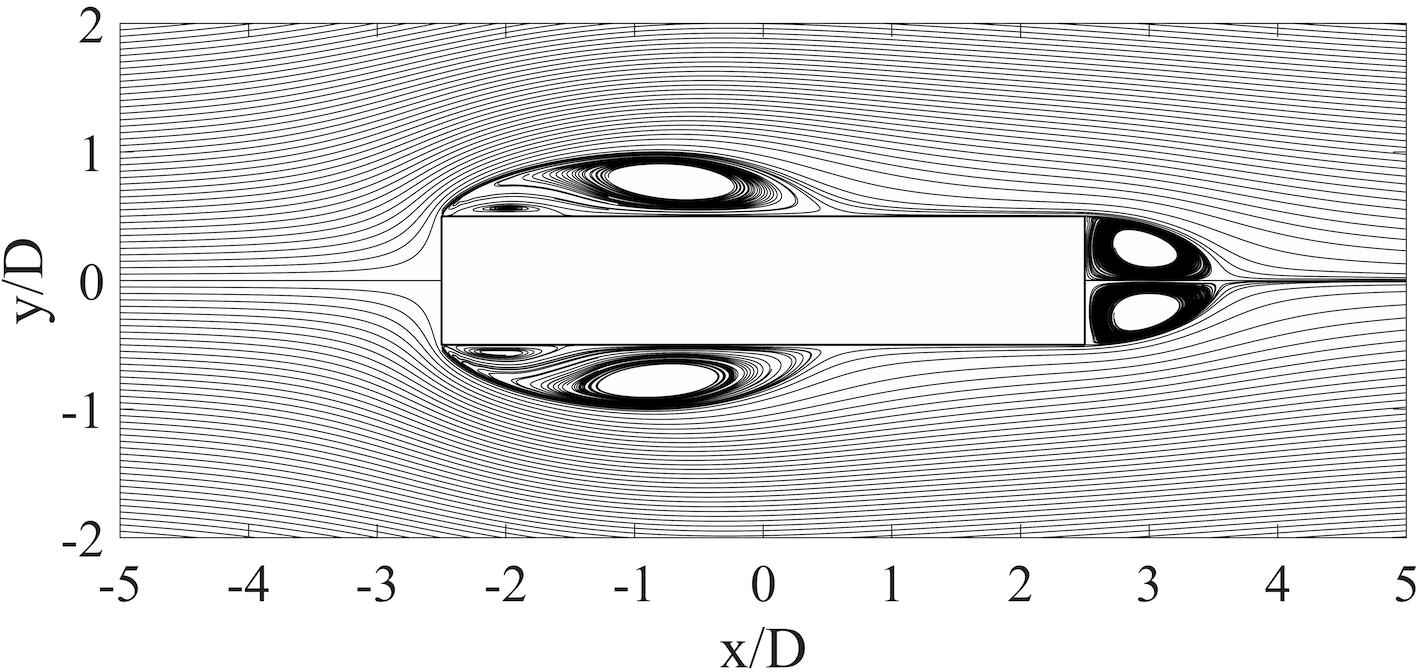}
		\includegraphics[width=.49\textwidth]{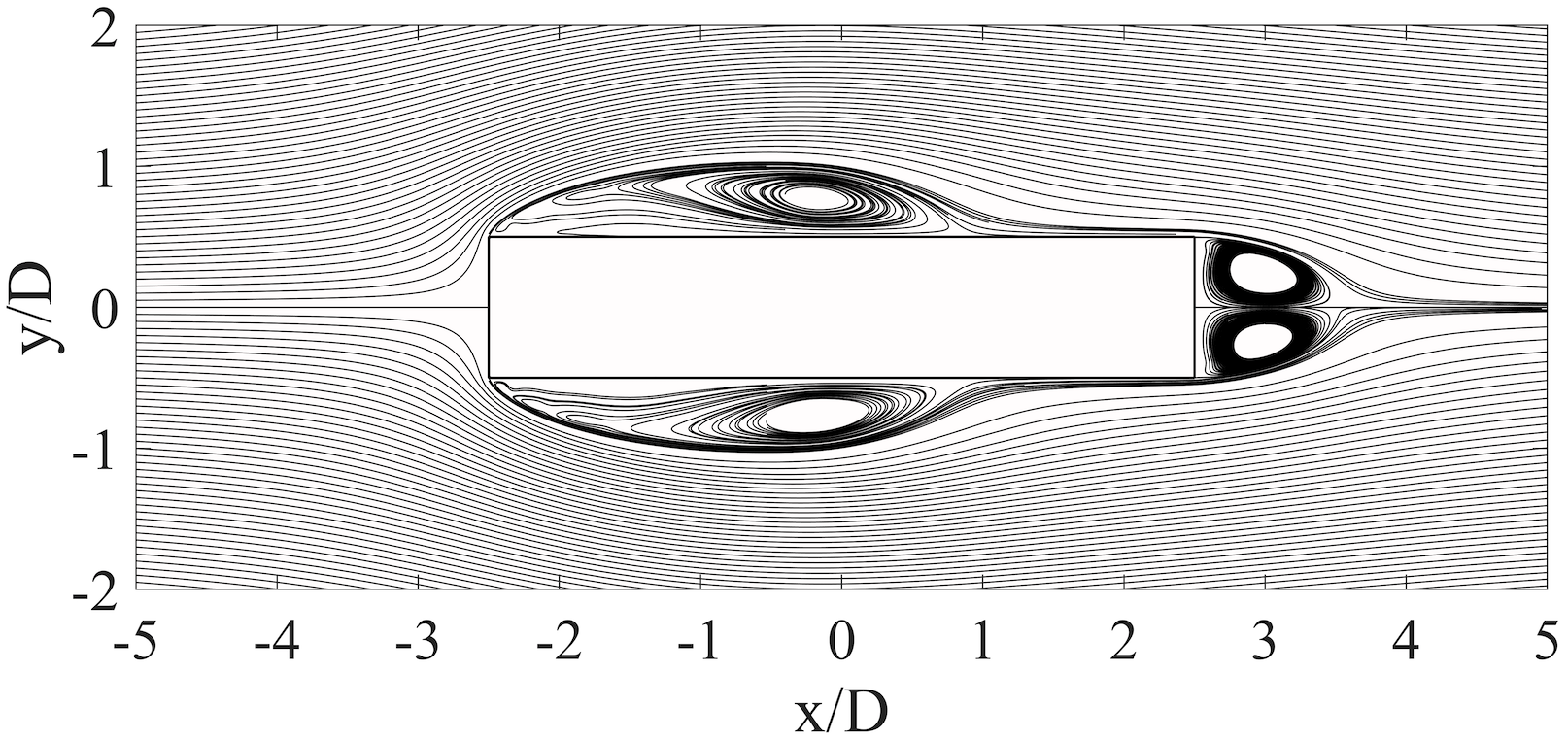}
		\hspace{0.4cm} (a) \hspace{6cm} (b)\\
		\caption{Time-averaged flow streamlines for (a) the \textit{reference LES} and (b) the \textit{baseline LES}.}
		\label{fig:ref_stream}
	\end{figure}

The mean flow streamlines for the \textit{reference LES} are shown in Fig.~\ref{fig:ref_stream}(a) and compared with those for the \textit{baseline LES} in Fig.~\ref{fig:ref_stream}(b). The velocity is averaged in time and in the spanwise direction. As expected for the low value of SGS dissipation adopted for the refined simulation, when compared with with the different results for the BARC benchmark, the numerical solution for the \textit{reference LES} is characterised by a short mean recirculation along the lateral side of the cylinder. As explained in Sec. \ref{sec:test_case}, the choice of a low value of SGS dissipation has the aim of maximising the difference between the results obtained with different grid resolutions to assess the proposed data assimilation procedure, regardless of the agreement with the experimental data. Indeed, the location of the mean reattachment point along the cylinder lateral side is at $x_r/D=0.58$, whereas the corresponding value for the \textit{baseline LES} is $x_r/D=0.89$. The value found in \cite{Rocchio2020} with the same grid resolution as the baseline LES but a larger SGS dissipation $w=0.05$, is $x_r/D=1.32$. This confirms that the low value of $w$ adopted in the present simulations tends to give shorter mean recirculation lengths.

\begin{figure}
		\centering
		\includegraphics[width=.65\textwidth]{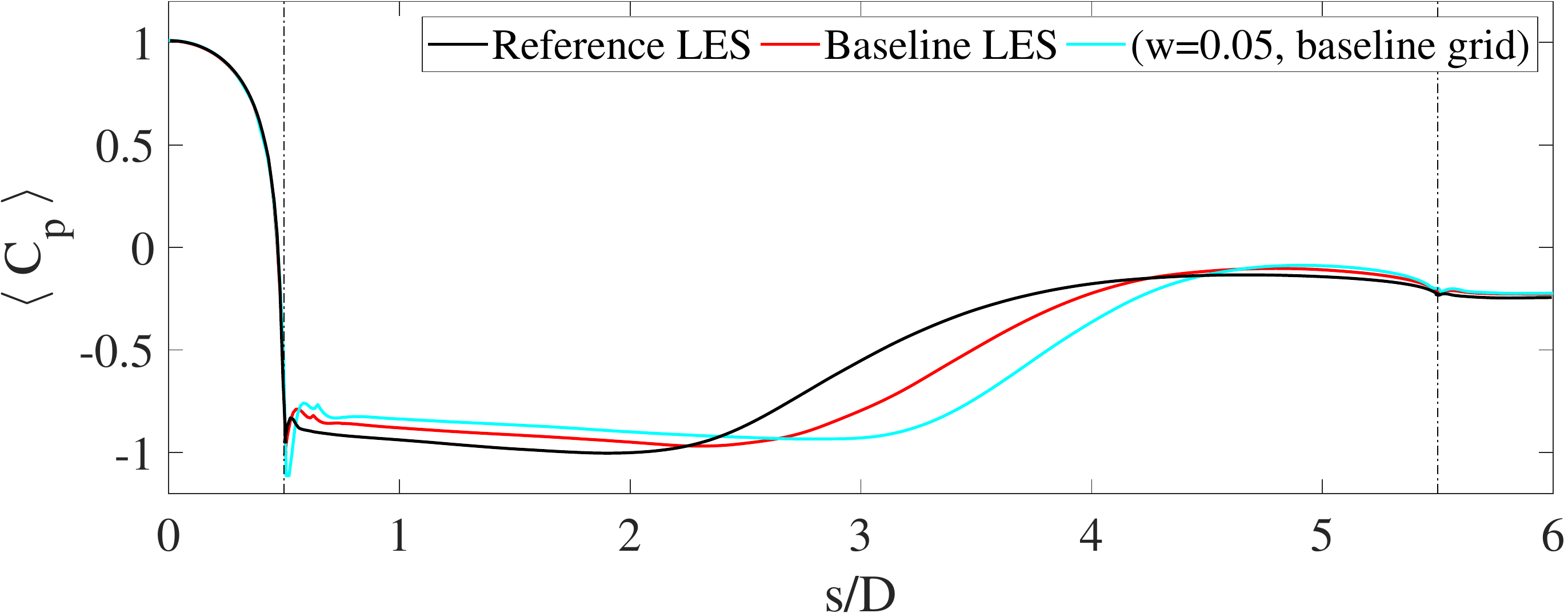}\\(a) \\
		\includegraphics[width=.65\textwidth]{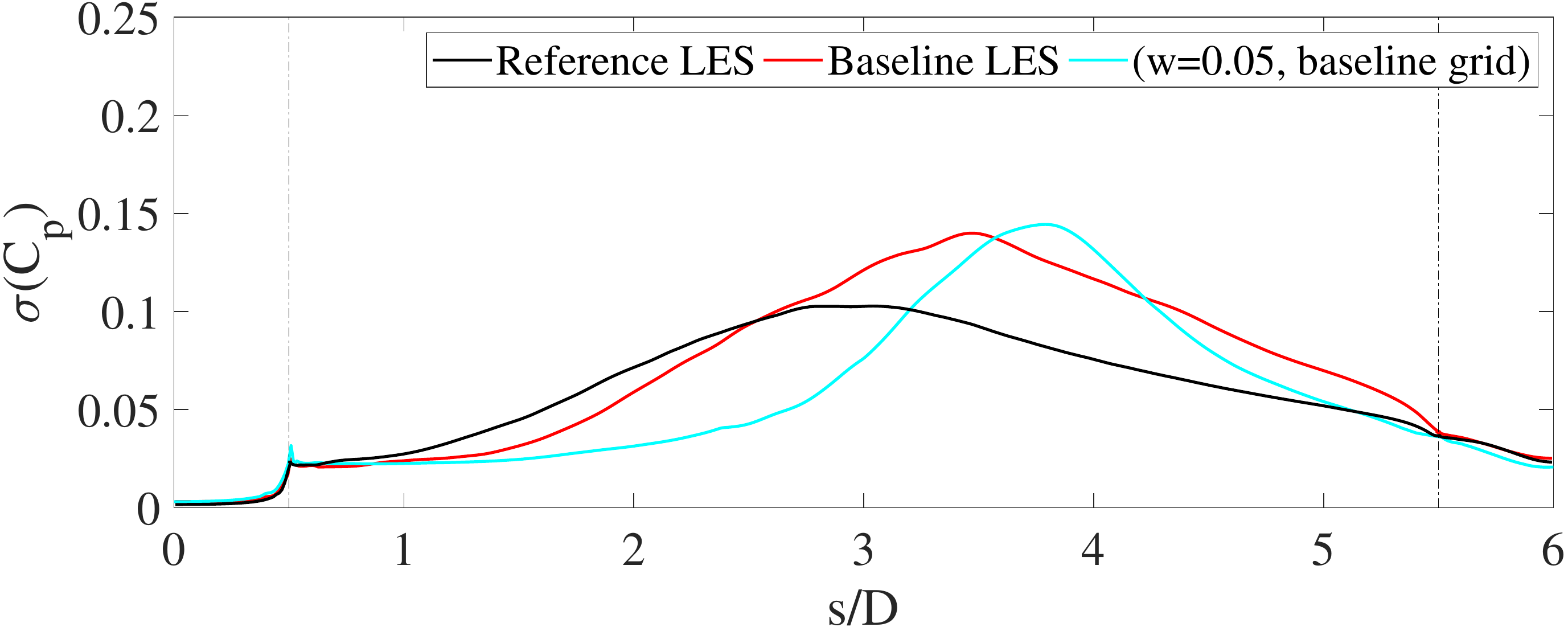}\\(b)
		\caption{(a) Pressure coefficient, $\langle C_p \rangle$, averaged in time, in the spanwise direction and between the upper and the lower surfaces of the cylinder. (b) Standard deviation of the pressure coefficient, $\sigma(C_p)$, averaged in space with the same criteria presented for $C_p$. The results for the \textit{reference LES} are compared with the \textit{baseline LES} and with the LES carried out with the baseline grid and $w=0.05$ in \cite{Rocchio2020}.}
		\label{fig:ref_p}
	\end{figure}

The pressure coefficient is defined as $C_p= \dfrac{p-p_\infty}{1/2 \rho U_\infty^2}$, where $p$ is the local pressure, $\rho$ is the density and $p_\infty$ is the freestream pressure. The distribution of the pressure coefficient averaged in time, in the spanwise direction and between the upper and the lower surfaces of the cylinder ($\langle C_p \rangle$) is shown in Fig.~\ref{fig:ref_p}(a). 
The local abscissa, $s/D$, is the distance from the cylinder stagnation point, which is located at $x/D=-2.5$ and $y/D=0$, measured along the cylinder side. 
The time standard deviation of the pressure coefficient, $\sigma(C_p)$, is reported in Fig.~\ref{fig:ref_p}(b). The distributions for the \textit{reference LES} are compared with the results for the \textit{baseline LES} and with the LES carried out in \cite{Rocchio2020} with the baseline grid and $w=0.05$. 
As shown in previous studies (see e.g. \cite{Bruno2014}), the location of the mean pressure recovery is related with the change of curvature of the mean streamlines occurring when the mean recirculation region starts to close. Similarly, the peak of $C_p$ fluctuations occur immediately downstream the mean flow reattachment \citep{Lunghi2022}.  Consistently with the mean flow pattern, thus, the $\langle C_p \rangle$ distribution obtained for the \textit{reference LES} is characterized by a pressure recovery that is located upstream the one for the \textit{baseline LES} and for the simulation carried out on the baseline grid with a larger value of $w$ by \cite{Rocchio2020}. Moreover, as expected, the location of the maximum value of $\sigma(C_p)$ is moved upstream for the \textit{reference LES}; the value of peak is consequently lower, consistently with the trend highlighted in \cite{Lunghi2022}.

\begin{figure}
		\centering
		\includegraphics[width=.49\textwidth]{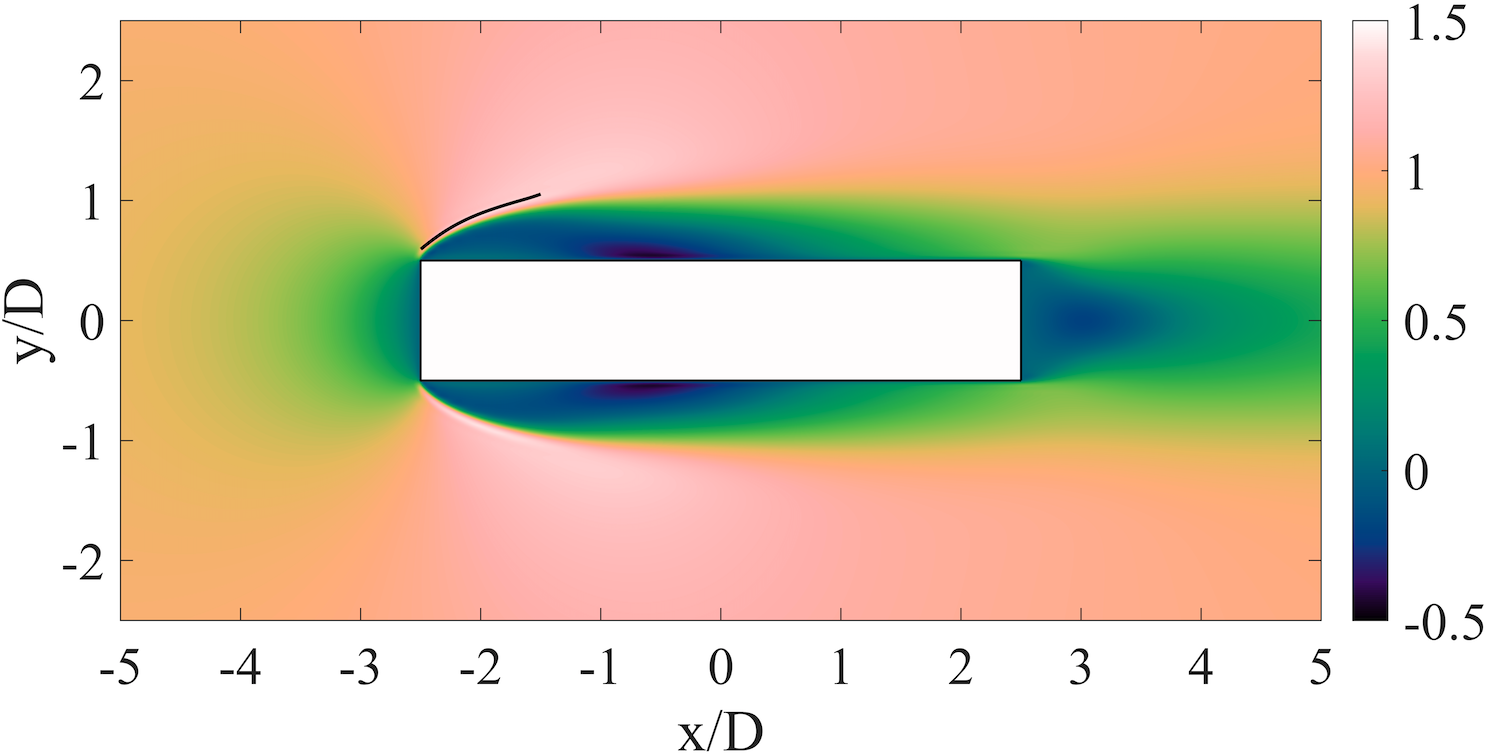}
		\includegraphics[width=.49\textwidth]{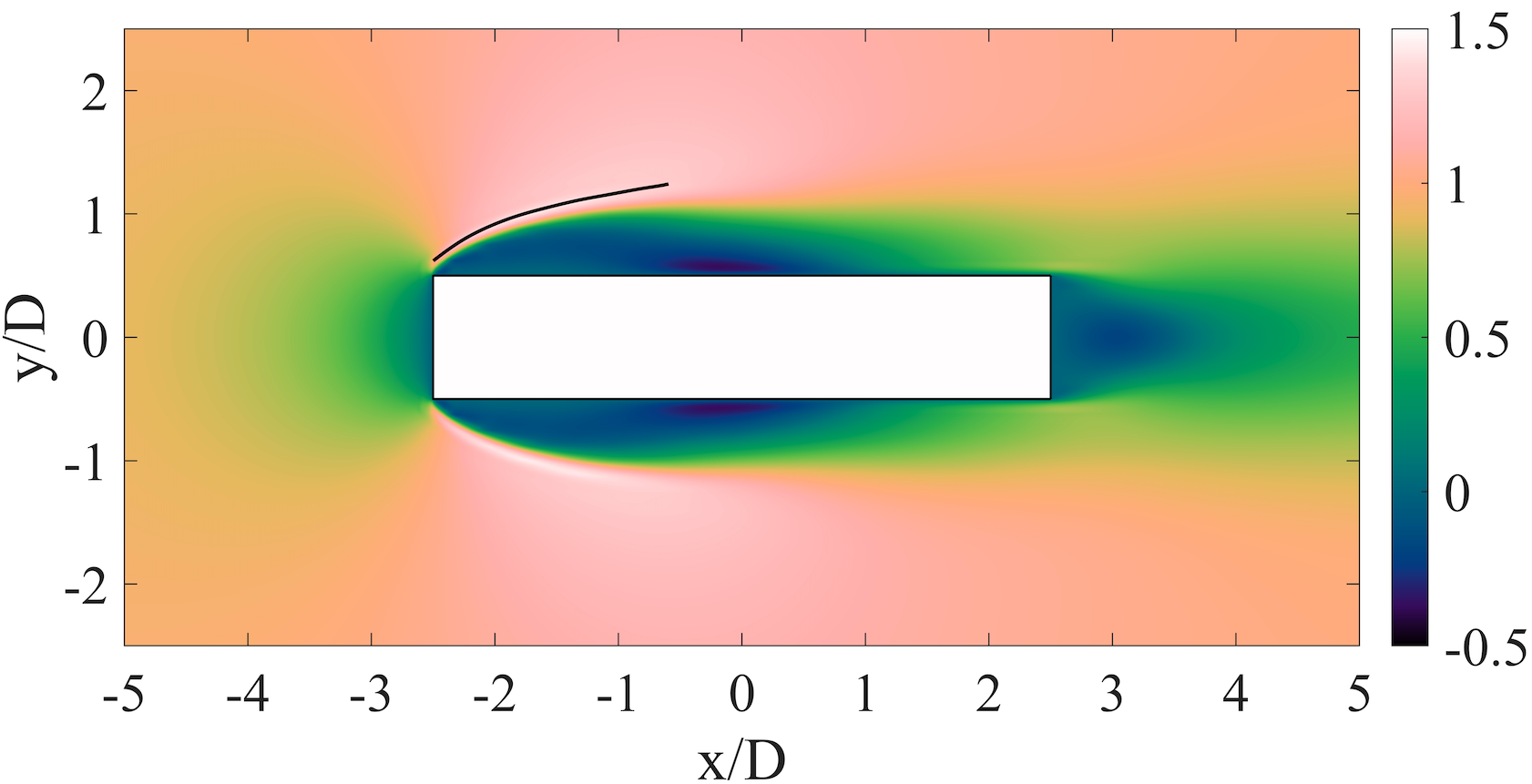}\\
		(a) \hspace{6cm} (b)\\
		\includegraphics[width=.47\textwidth]{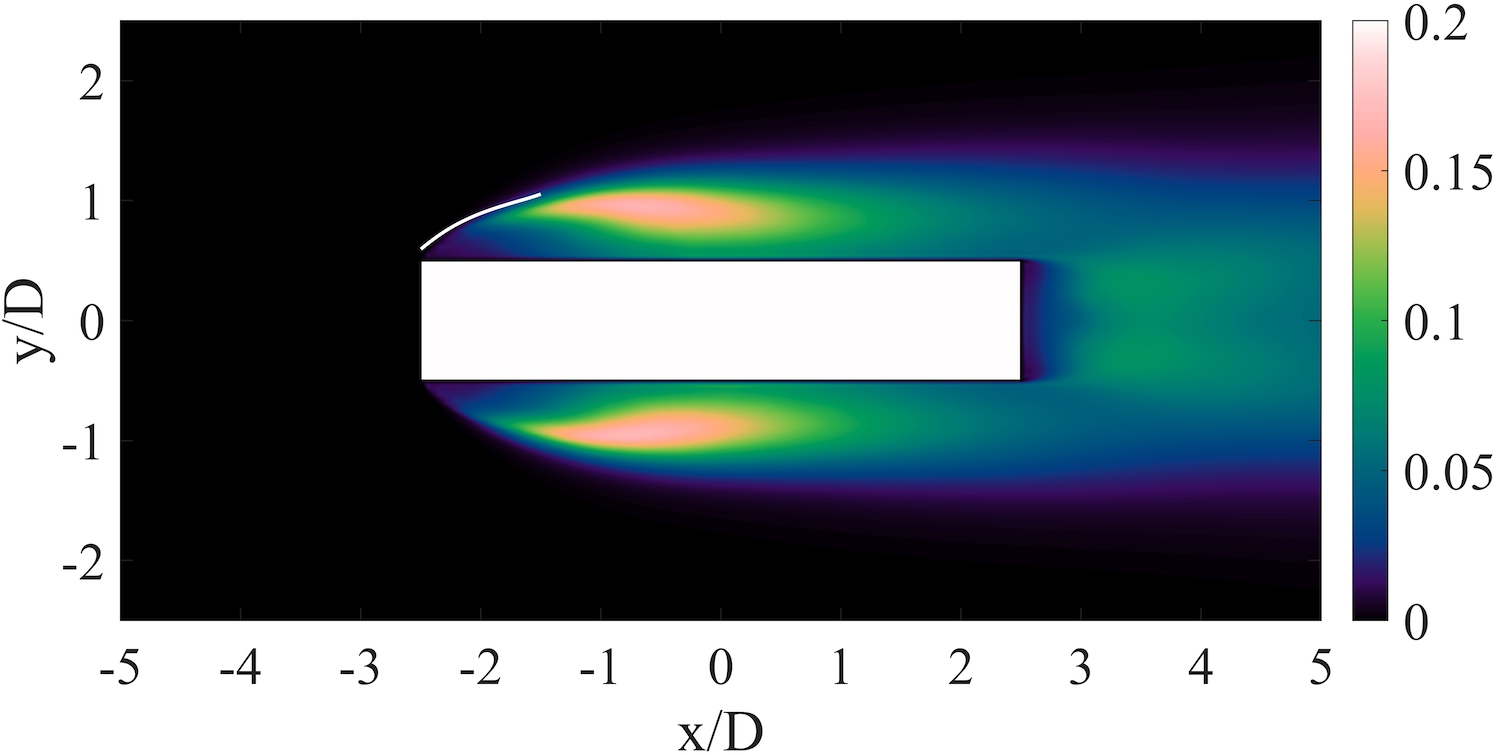}
		\includegraphics[width=.51\textwidth]{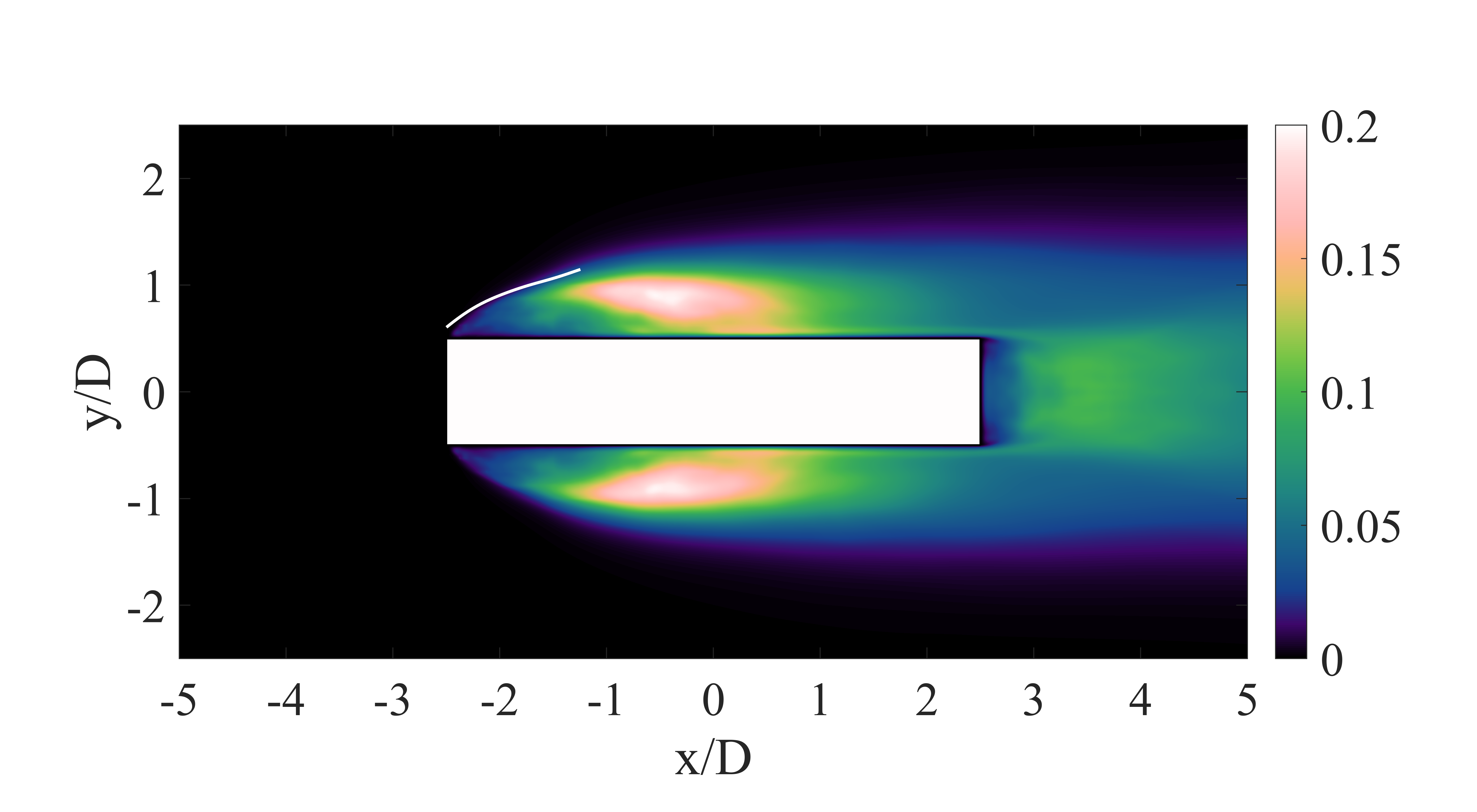}\\
		(c) \hspace{6cm} (d)\\
		\caption{(a,b) Mean streamwise velocity, $\langle u \rangle$, and (c,d) turbulent kinetic energy, $\textup{TKE}$. Results for (a,c) the \textit{reference LES} and (b,d) the \textit{baseline LES}. The lines represent the positions of the outer border of the detaching shear layers.}
		\label{fig:ref_field}
	\end{figure}

\begin{figure}
		\centering
		\includegraphics[width=.65\textwidth]{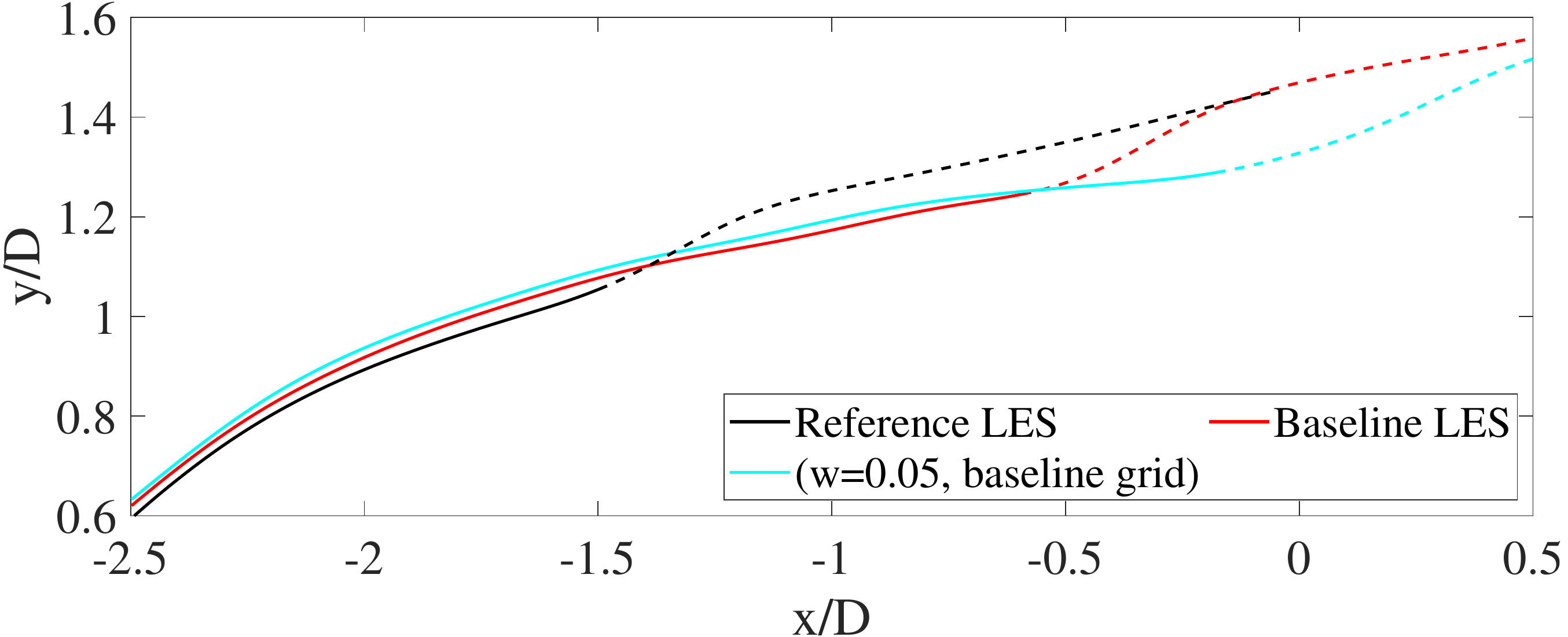}\\(a)\\
		\includegraphics[width=.65\textwidth]{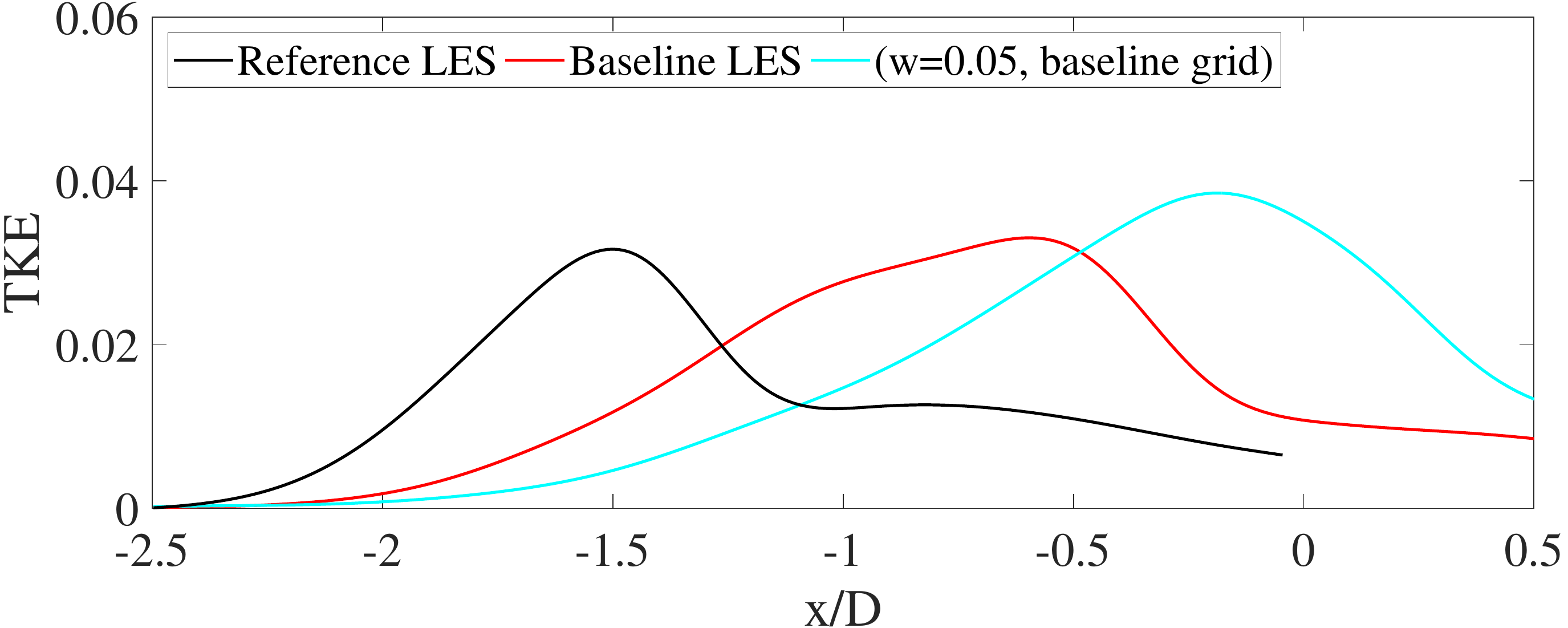}\\(b)\\
		\includegraphics[width=.65\textwidth]{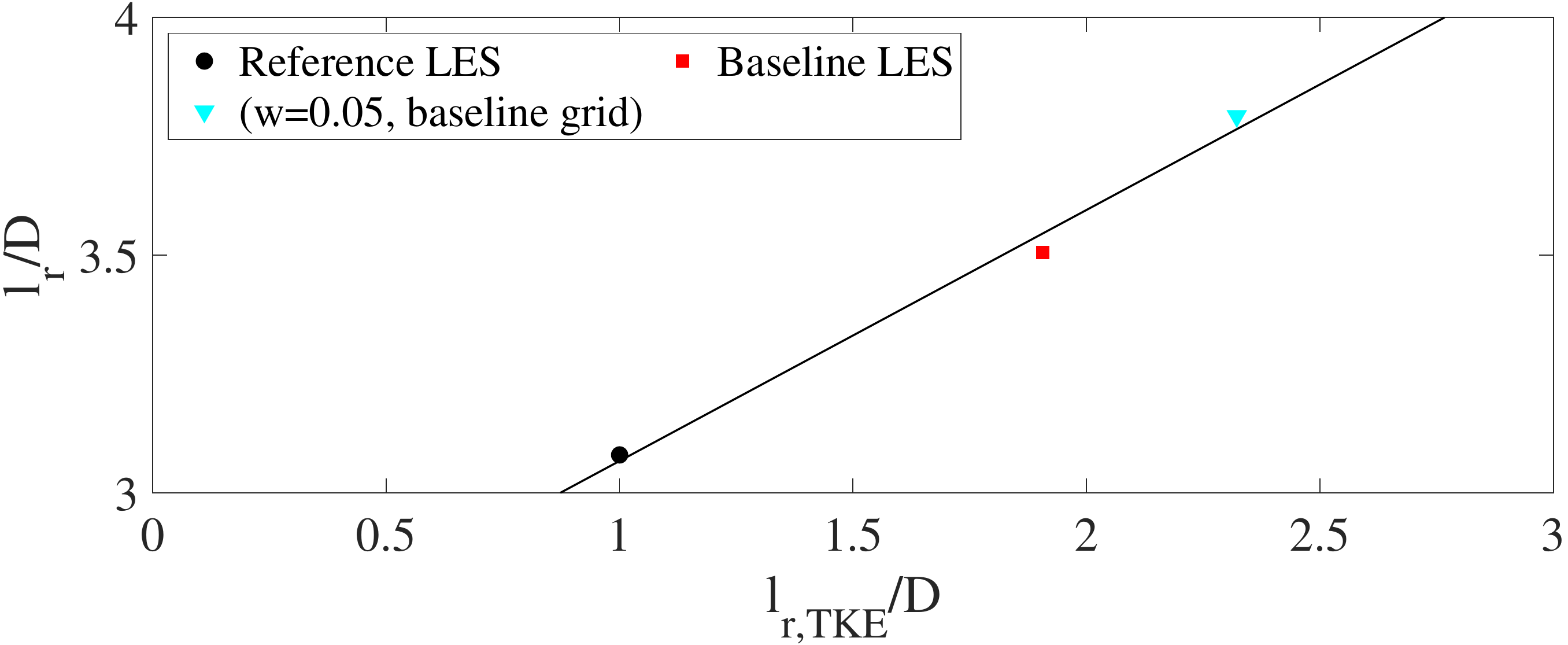}\\(c)
		\caption{(a) Position of the outer border of the detaching shear layer, (b) distribution of the $\textup{TKE}$ along the outer border of the detaching shear layer, and (c) correlation between the length of the mean recirculation region, $l_r$, and the position of maximum TKE on the shear layer border evaluated from the upstream corner, $l_{\textup{r,TKE}}$. The results for the \textit{reference LES} are compared with the ones for the \textit{baseline LES} and with the LES carried out with the baseline grid and $w=0.05$ in \cite{Rocchio2020}.}
		\label{fig:ref_border}
	\end{figure}

\begin{figure}
		\centering
		\includegraphics[width=.45\textwidth]{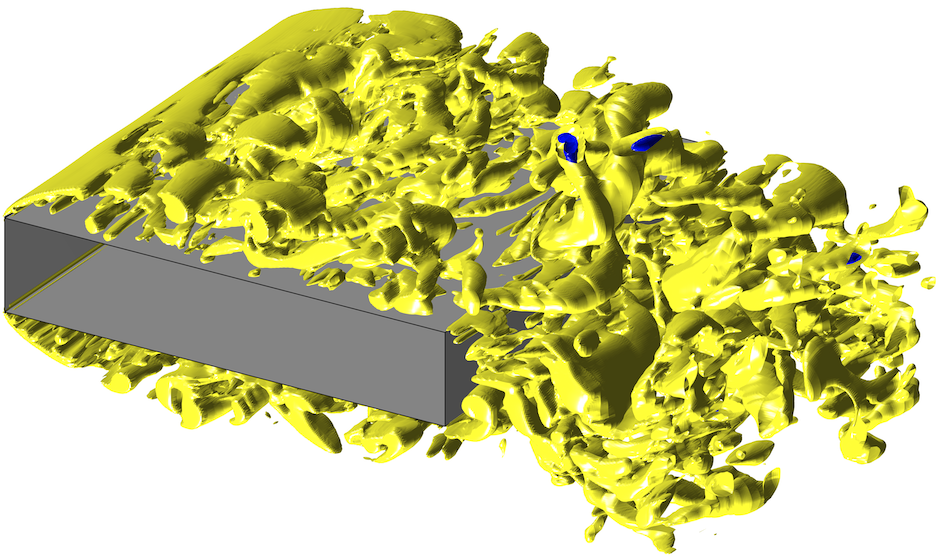}
		\includegraphics[width=.45\textwidth]{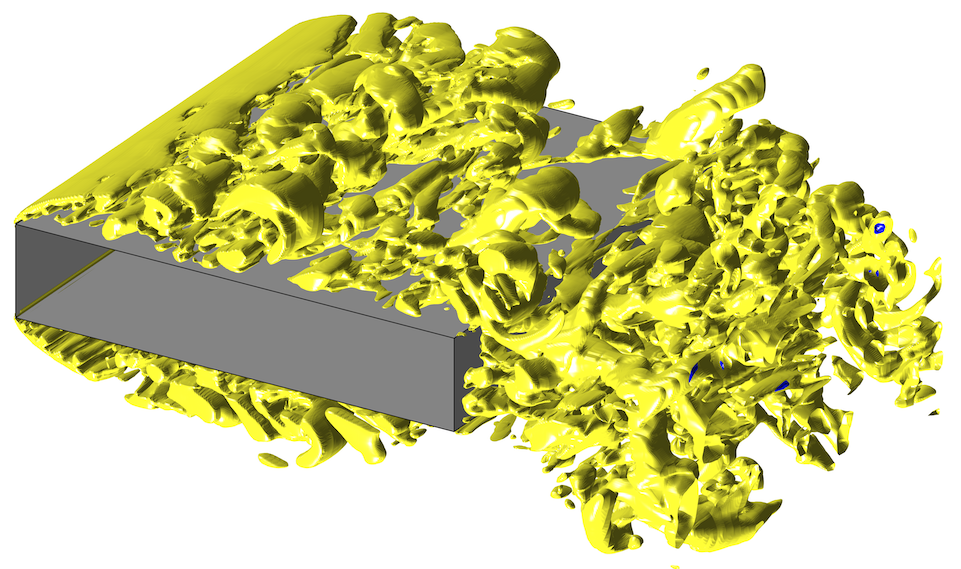}\\
		(a) \hspace{6cm} (b)\\
		\includegraphics[width=.47\textwidth]{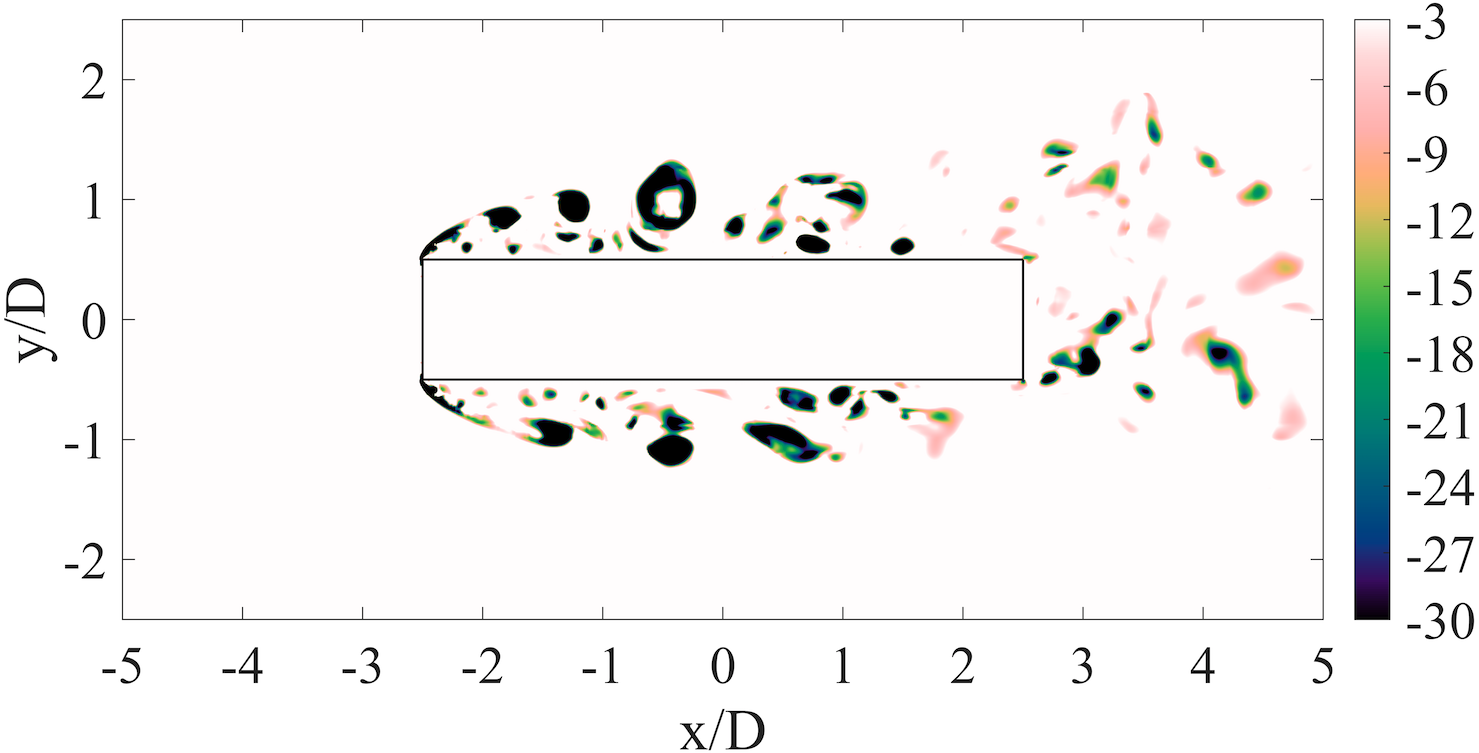}
		\includegraphics[width=.52\textwidth]{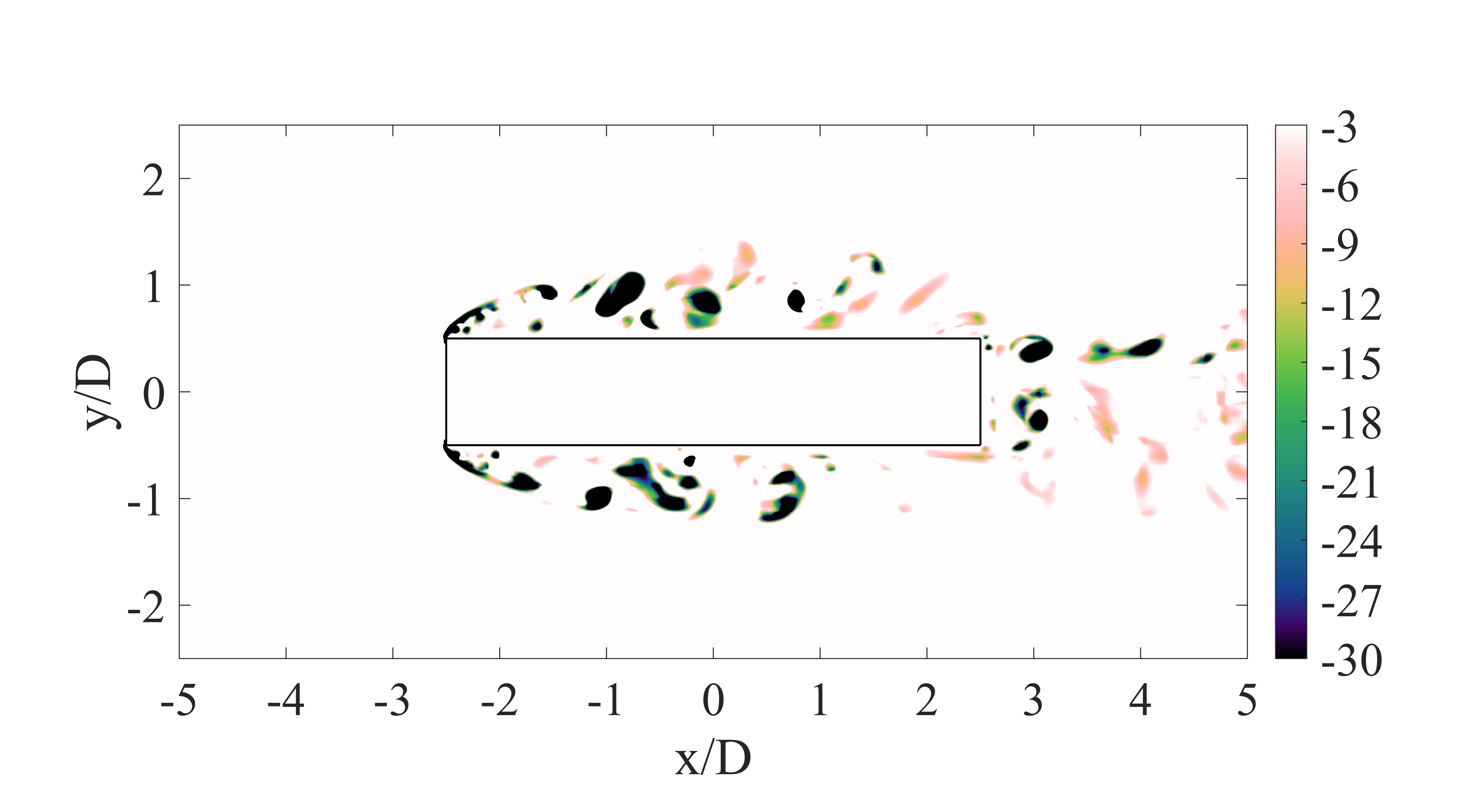}\\
		(c) \hspace{6cm} (d)\\
		\caption{(a,b) Isocontours of the instantaneous vortex indicator $\lambda_2$ and (c,d) its distribution on the spanwise symmetry plane ($z/D=0$). Results for (a,c) the \textit{reference LES} and (b,d) the \textit{baseline LES}.}
		\label{fig:ref_l2}
	\end{figure}

The mean streamwise velocity, $\langle u \rangle$, and the turbulent kinetic energy, $\textup{TKE}=1/2(\sigma^2(u)+\sigma^2(v)+\sigma^2(w))$ fields for the \textit{reference LES} are shown in Figs.~\ref{fig:ref_field}(a) and (c), respectively. The results for the \textit{baseline LES} are provided in Figs.~\ref{fig:ref_field}(b) and (d). The solid line starting just downstream the upper upwind corner sketches the position of the outer border of the detaching shear layer, evaluated from the averaged velocity field following the procedure in \cite{Rocchio2020}. It can be appreciated how the TKE increases moving downstream along the shear-layer, although the maximum in the field is located in the rear part of the recirculation region. 
Previous studies \citep{Rocchio2020,Lunghi2022} indicated that  there is an almost perfect correlation between the length of the mean recirculation region and the location at which the maximum of TKE is reached along the border of the detaching shear layer: the more usptream the peak of TKE occurs the shorter is the mean recirculation length. Figures~\ref{fig:ref_border}(a) and (b) compare the position of the outer border of the detaching shear layer and the distribution of $\textup{TKE}$ along this border between the \textit{reference LES}, the \textit{baseline LES} and one LES carried out with the baseline grid and $w=0.05$ in \cite{Rocchio2020}.
The position of the outer border of the shear layer detaching from the upstream corner almost coincides in the three considered cases, but a more upstream position of maximum TKE, $l_{\textup{r,TKE}}$, is found for the \textit{reference LES}. This is consistent with the previously described  correlation between the length of the mean recirculation region and the position of maximum TKE on the shear layer border and it is also quantitatively confirmed in Fig.~\ref{fig:ref_border}(c). From a physical point of view, the streamwise extension of the mean recirculation region is related with the location of the Kelvin-Helmholtz instability onset in the detaching shear-layers, and this is in turn correlated with the position of the maximum kinetic energy along the shear-layer edge. The visualisation of the isocontours of the instantaneous vortex indicator $\lambda_2$ in Fig.~\ref{fig:ref_l2} show this physical mechanism of instability and roll-up of the detaching shear-layer. In accordance with all the previous considerations, it can be observed that the  roll-up of the shear layers for the \textit{reference LES} occurs upstream than for the \textit{baseline LES}.

One could argue that the set-up chosen for the \textit{refined LES}, in terms of grid and value of the parameter $w$, produces a short recirculation bubble when compared with experimental observation. As a consequence, also the main statistical quantities exhibit marked differences. This apparently incoherent choice has been performed because of the enhanced sensitivity of LES producing short recirculation bubbles to variations in the mesh resolution. Therefore, this set up will act as a magnifying lens over the performance of the MGEnKF, allowing for an unambiguous assessment of the LES optimization procedure. 	
	
	
\section{Data-driven enhancement of LES using the MGEnKF}
\label{sec:DA-LES}
\subsection{Description of the DA experiment}
The MGEnKF method is now used to augment the predictive capabilities of LES on coarse grids. The resulting data-driven simulation, which will be referred to as \textit{DA augmented LES} (see Table~\ref{tab:setupLES}), corresponds to a cluster of calculations in which the main simulation, which uses the baseline grid (the same grid used for the \textit{baseline LES}), is assisted by $N_e$ LES, which are run on the \textit{coarse grid}. The state estimation and the parametric inference steps are performed via this ensemble of LES.

\begin{figure}
	\centering
	\includegraphics[width=1\textwidth]{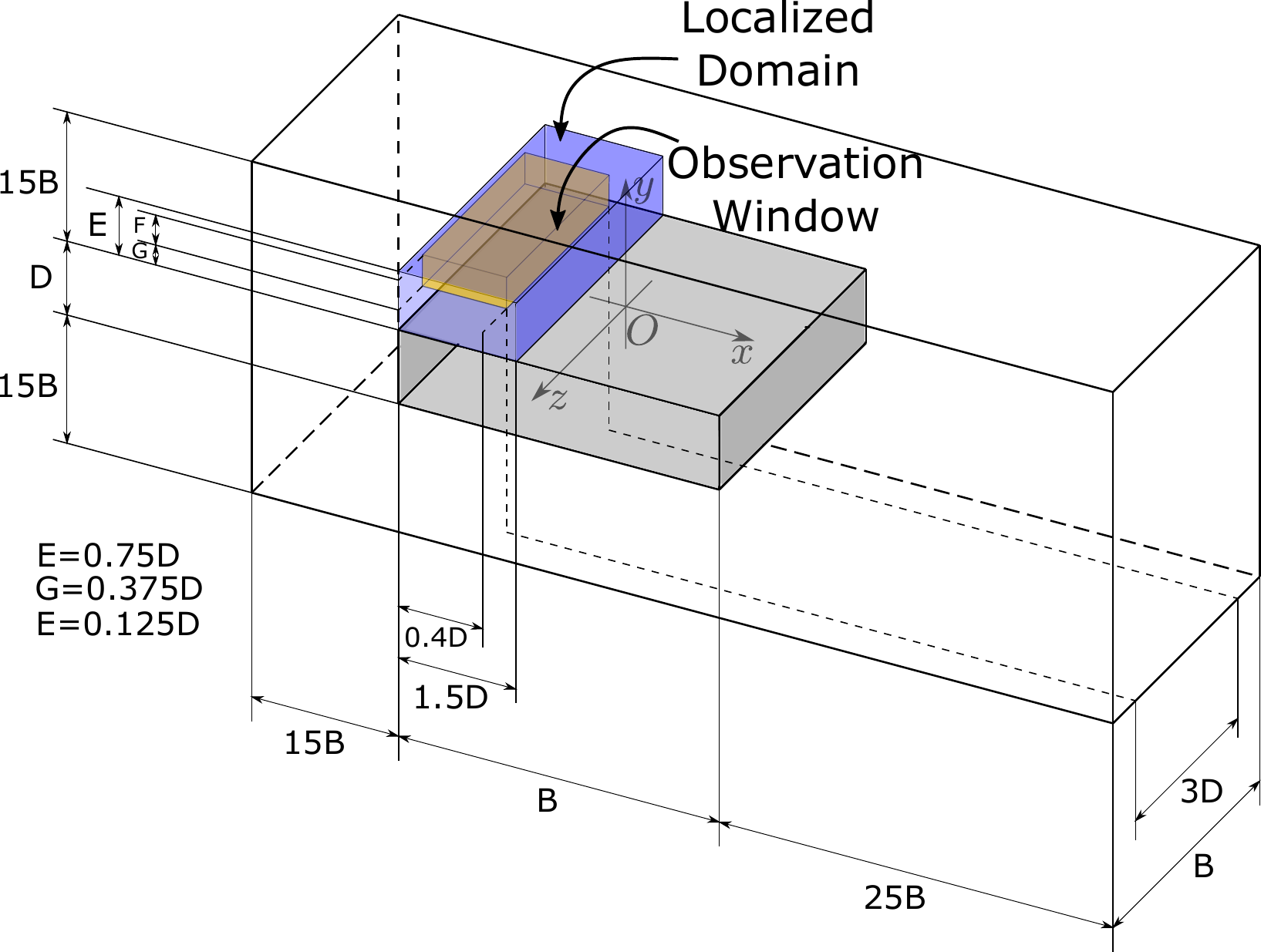}\\
	\caption{Representation of the physical domain around the rectangular cylinder, represented in grey. The physical region where sensors are placed is indicated in yellow, while the domain of application of the Kalman update via localization is indicated in blue.}
	\label{fig:DA_domain}
\end{figure}

Details are now provided for each ingredient needed for the application of the Data Assimilation algorithm. Observation is obtained by sampling the three components of the instantaneous velocity field of the \textit{reference LES}, which is run for a total time of $T=200t_c$, where $t_c=D/U$ is the characteristic advection time. A cloud of sensors is embedded in the region $-2.25\leq x/D \leq -1.75$, $0.75\leq y/D \leq 1$, $1\leq z/D \leq 4$ (see Fig.~\ref{fig:DA_domain}) 
and data is sampled every $0.1t_c$. It is considered that observation is available on all the points of the coarse grid inside the observation region. 
Roughly, 4000 local snapshots of the instantaneous flow are captured by the sensors. In order to increase the degree of randomization of the sampled data, artificial Gaussian noise of covariance $R=\sigma_o^2I$ is applied to the observation, where $I$ is the identity matrix. Here, $\sigma_o$ is set to $0.02$, and the artificial noise added is considered to be uncorrelated in time, therefore $R$ is diagonal. 
This noise, tailored to mimic realistic experimental measurements, improves the rate of convergence of EnKF-based tools, as shown in several research works \citep{Tandeo_Ailliot_Bocquet_Carrassi_Miyoshi_Pulido_Zhen_MWR_2020,Moldovan2021_jcp}. One should also take into account that the sampling strategy of the observation intentionally lacks symmetry, as it could be expected in realistic operative conditions. This important point will be extensively discussed in the following. 
	
\begin{figure}
	\centering
	\includegraphics[width=0.8\textwidth]{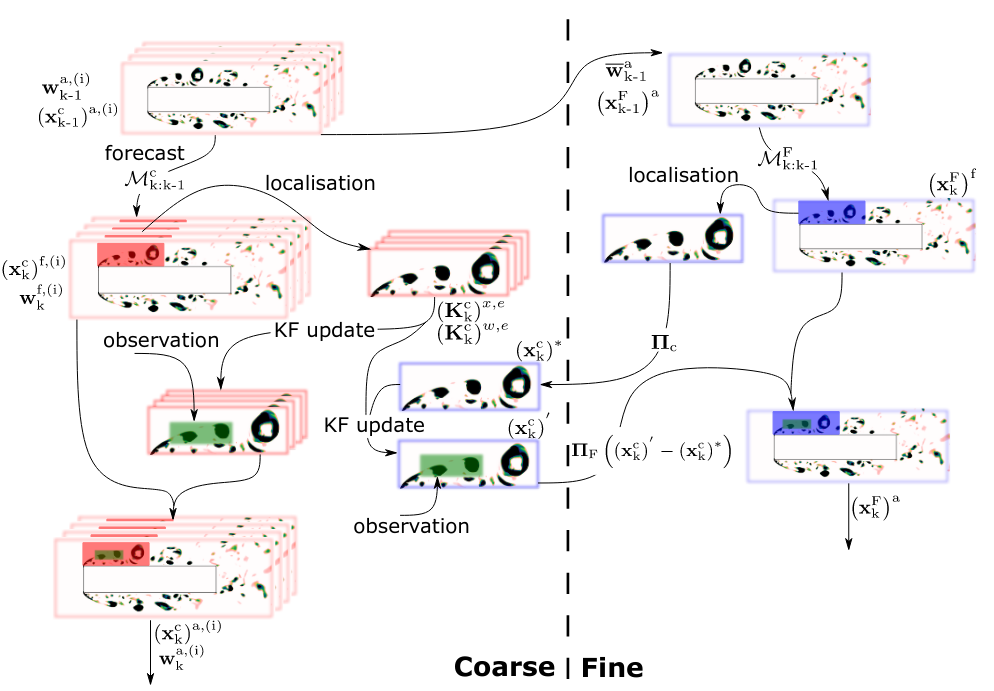}\\
\caption{Application of the MGEnKF algorithm to the BARC test case.}
	\label{fig:coupling_schema}
\end{figure}

The code Nek5000 is embedded within the MGEnKF algorithm with the help of an opensource coupler (OpenPALM) \citep{Duchaine_2015}. As previously discussed, the \textit{DA augmented LES} runs in a multilevel setting. The main simulation is run on the baseline grid, while the ensemble members are run on the \textit{coarse grid}. A scheme of the application of the MGEnKF to this specific test case is shown in Fig.~\ref{fig:coupling_schema}. Details about the computational set-up are provided in Table \ref{tab:setupLES}. One can see that the grid used for the calculations of the ensemble members is around two times coarser than the baseline grid and four times coarser than the refined grid in each space direction.

The \textit{DA augmented LES} is performed in the following conditions:

\begin{itemize}
    \item The initial solution imposed to the main and ensemble simulations is obtained from the reference simulation by interpolating the first time step solution on the baseline and coarse grids. This choice allows avoiding synchronization issues between observation and model prediction. While this is clearly an interesting topic which has been investigated in recent studies \citep{wang_zaki_2022}, we decided to restrict the present work to the analysis of the performance of the MGEnKF in controlled conditions. 
    \item Considering that the outer loop of the MGEnKF works exclusively using the ensemble members, the observation sampled from the reference simulation is directly projected on the collocation points of the coarse grid for sake of simplicity. For this case, the observation is interpolated on a total of $6$ spectral elements. The size of the observation vector is thus approximately $4\cdot10^3$. It is important to stress here that, unlike classical KF approaches, the most complex manipulation in EnKF is a matrix inversion whose size is the same of the observation vector.
    \item The value of the ensemble size $N_e$ is determined as a compromise between computational cost and a good representation of the solution. We set here $N_e=30$.
    \item Initial values of $w$ for each member of the ensemble $w^{(i)}\,,i=1,2,3...,N_e$ are sampled from the Gaussian distribution $\mathcal{N} \left(w_0,R_w \right)$, where $w_0$ is the prior value $w_0=0.018$ and $R_w=0.0001^2 I$. We recall that $w$ regulates the amount of introduced dissipation in the spectral space, which may be considered as a SGS dissipation. As discussed previously, and as shown in \cite{Mariotti2017}, this parameter as a key impact on the LES results, together with grid resolution.
    \item Observation is sequentially integrated into the system to: i) optimize the ensemble of parameters $w^{(i)}$, ii) correct the physical state of each member of the ensemble. The DA state update is extended towards the main simulation owing to interpolation functions which connect the coarse and baseline grids. In addition, the parametric description of main simulation is also optimized updating the parameter value with $\mathbb{E}\left[w\right]$, i.e. $\overline{w^{(i)}}$.
\end{itemize}
\subsection{Simplifications applied to the MGEnKF}
A few simplifications have been applied to the classical MGEnKF algorithm to make it more suitable to this HPC application. First, the state vector includes the three components of the velocity field but not the pressure, which is updated via a Poisson equation at the beginning of the following time step. The nature of the Poisson equation allows to rapidly propagate information provided in the DA update \citep{Meldi2017_jcp}. Therefore, the MGEnKF does not grant conservativity at the end of the analysis phase, but the momentum and mass conservation are satisfied at the following time step. This simplification allows reducing sensibly the size of the state vector without producing numerical instabilities in the analysis phase, owing to the effect of the Poisson equation. In addition, the effect of the Kalman gain has been \emph{localized}.
In practice, a correlation matrix $\mathbf{B}$ is introduced in \eqref{eq:ensemble_update} as follows \citep{SOARES2018110}:

\begin{equation}
\mathbf{x}_k^{\text{a},(i)}=
\mathbf{x}_k^{\text{f},(i)}+
\mathbf{B}\odot\mathbf{K}_k^\text{e}
\left(y_k^{\text{o},(i)}-\mathbf{\mathcal{H}}_k\left(\mathbf{x}^{\text{f},(i)}_k\right)\right),\label{eq:ensemble_update_localized}
\end{equation}

where $\odot$ is the Schur product (element-wise product between $\mathbf{B}$ and $\mathbf{K}_k^\text{e}$). $\mathbf{B}$ can be estimated as a distance-dependent correlation matrix constructed as 

\begin{equation}\label{beta}
    B_{ij}=\exp \left[-\frac{d\left(i,j \right)^2}{2L^2}\right],
\end{equation}
see \citet{Gaspari_1999}, where $d\left(i,j \right)$ represents the euclidean distance between a point $i$, where observation is available, and a model grid point $j$, and $L$ is the correlation distance ($L=1D$ in our case).
This implies that the matrix $\mathbf{K}_k^\text{e}$ is transformed using a spatial Gaussian filter. The product between the correlation matrix $\mathbf{B}$ and the Kalman gain erases any spurious long-range statistical correlations (beyond a length $L$) that might appear. 
In addition, all the velocity updates via Kalman filter have been set to zero outside a prescribed domain defined by the coordinates: $-2.5\leq x/D \leq -1$, $0.5\leq y/D \leq 1.25$, $0\leq z/D \leq 5$. This strategy is used to avoid spurious correlation to affect the precision of the results, as well as to further reduce the computational costs. Using this strategy, the size of the \emph{localized} state vector updated via data assimilation is approximately $6\cdot10^4$, around $30$ times less than the complete state vector on the coarse grid.

As previously discussed, the first version of the MGEnKF presented in \cite{Moldovan2021_jcp} is here used, and it has been preferred to the most recent version presented in \cite{Moldovan2022}. The difference between the two versions consists in the usage of an \textit{inner} loop for the latter, which further improves the accuracy of the model used for the coarse grid level. However, the efficiency of the \textit{inner} loop has been verified for 1D linear and non-linear test cases, and application to complex 3D cases was deemed too ambitious for the present work. Therefore, it was decided to devote the present work to the analysis of a complex case using only \textit{outer loop}, which represents in any case the essential element of the reduced-order EnKF procedure.

\subsection{Results}
Starting from the initial solution, the coarse grid ensemble members and the main simulation are sequentially augmented at each analysis phase. This includes an optimization of the parameter $w$ as well as a state update for every simulation (main and ensemble members).

\begin{figure}
	\centering
	{\includegraphics[width=0.75\textwidth]{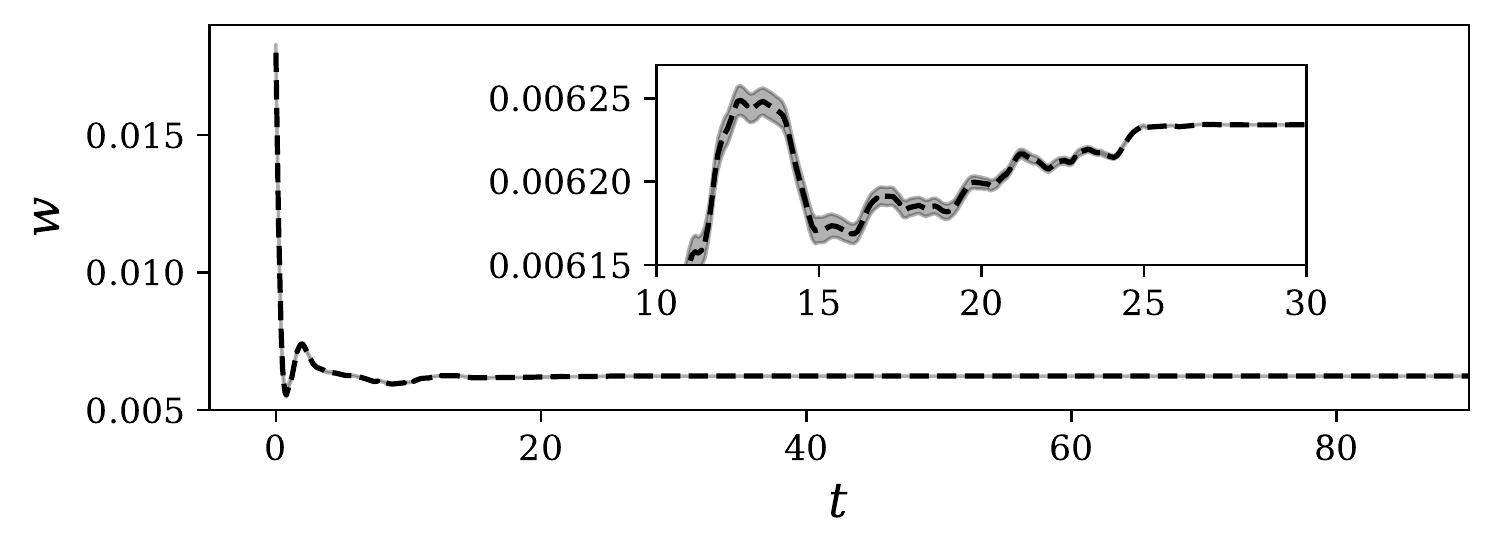}}\\
	\caption{Convergence of the MGEnKF optimization of the explicit LES coefficient filter $w$ applied to the resolved equations. The coefficient is initialized with the baseline value. 
	The shaded area represents the $95\%$ confidence interval.}
	\label{fig:w}
\end{figure}

The results obtained via MGEnKF are now analysed. Figure~\ref{fig:w} shows the time evolution of $w$, which is sequentially optimized during the analysis phases. One can see that the value of $w$ rapidly moves away from the prior ($w=0.018$). It first exhibits some fluctuations before converging to $w=0.00623$ in approximately $25$ advection time units. A progressive reduction of the variance in the estimation (shaded area) is also observed, which is a solid indicator of convergence.  The optimized value exhibits a large difference when compared with the prior ($0.018$ vs $0.00623$). An independent LES simulation without DA, referred to as \textit{optimized LES}, has been run on the baseline grid using the optimized value of $w=0.00623$. 
In order to provide a complete picture of the effectiveness of the method, the four simulations described in Table \ref{tab:setupLES} are compared in the following: the \textit{reference LES}, the \textit{baseline LES} ($w=0.018$), the \textit{data-augmented LES}, and the \textit{optimized LES} ($w=0.00623$). We stress that the MGEnKF is applied only on the \textit{data-augmented LES} simulation.

\begin{figure}
		\centering
		\includegraphics[width=.49\textwidth]{refined_streamlines}
		\includegraphics[width=.49\textwidth]{baseline_streamlines}\\
		\hspace{0.4cm} (a) \hspace{6cm} (b)\\		\includegraphics[width=.49\textwidth]{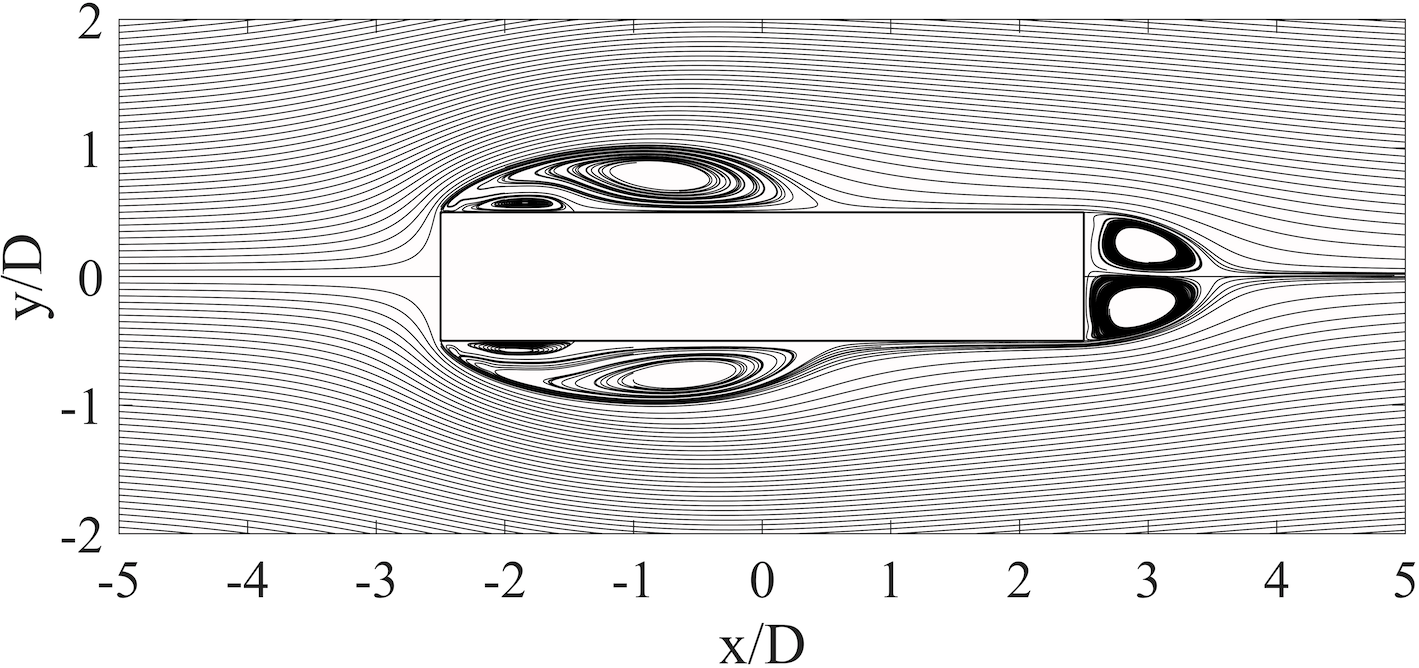}
		\includegraphics[width=.49\textwidth]{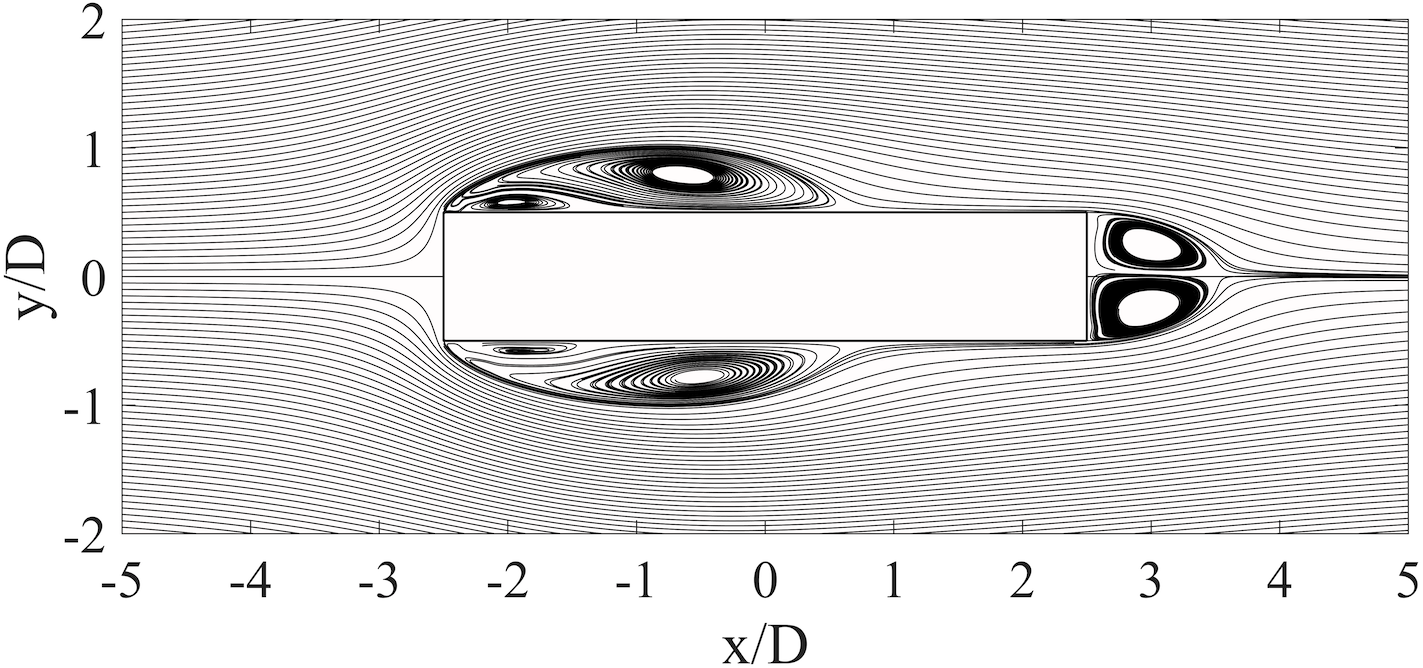}\\
		\hspace{0.4cm} (c) \hspace{6cm} (d)\\
		\caption{Time-averaged flow streamlines for (a) the \textit{reference LES}, (b) the \textit{baseline LES}, (c) the \textit{data-augmented LES} and (d) the \textit{optimised LES}}
		\label{fig:DA_stream}
	\end{figure}

The time-averaged flow streamlines provided for the four simulations in Fig.~\ref{fig:DA_stream} seem to indicate that the size of the recirculation regions for the \textit{DA augmented LES} and the \textit{optimized LES} is significantly closer to the prediction obtained via the \textit{reference LES}, when compared with the prior \textit{baseline LES}. One striking result that can be observed in this figure is that, despite the fact that the observation is provided only in the region on top of the domain, the average flow field predicted by the \textit{DA augmented LES} is convincingly symmetric and the improvement in the accuracy is not just localized to the observation region. This global accuracy improvement despite the sparsity of the observation is due to the relative complexity of the CFD model and probably to the effect of the Poisson equation \citep{Meldi2017_jcp}. In addition, the optimization of the global parameter $w$ which affects the entire computational domain, may also be play an important role in obtaining symmetrical flow statistics. Therefore, present results seem to indicate that coupling data-driven tools with relatively complex physical models, such as CFD, may prove efficient even with limited observation, which is the case for numerous practical applications.

\begin{figure}
		\centering
		\includegraphics[width=.65\textwidth]{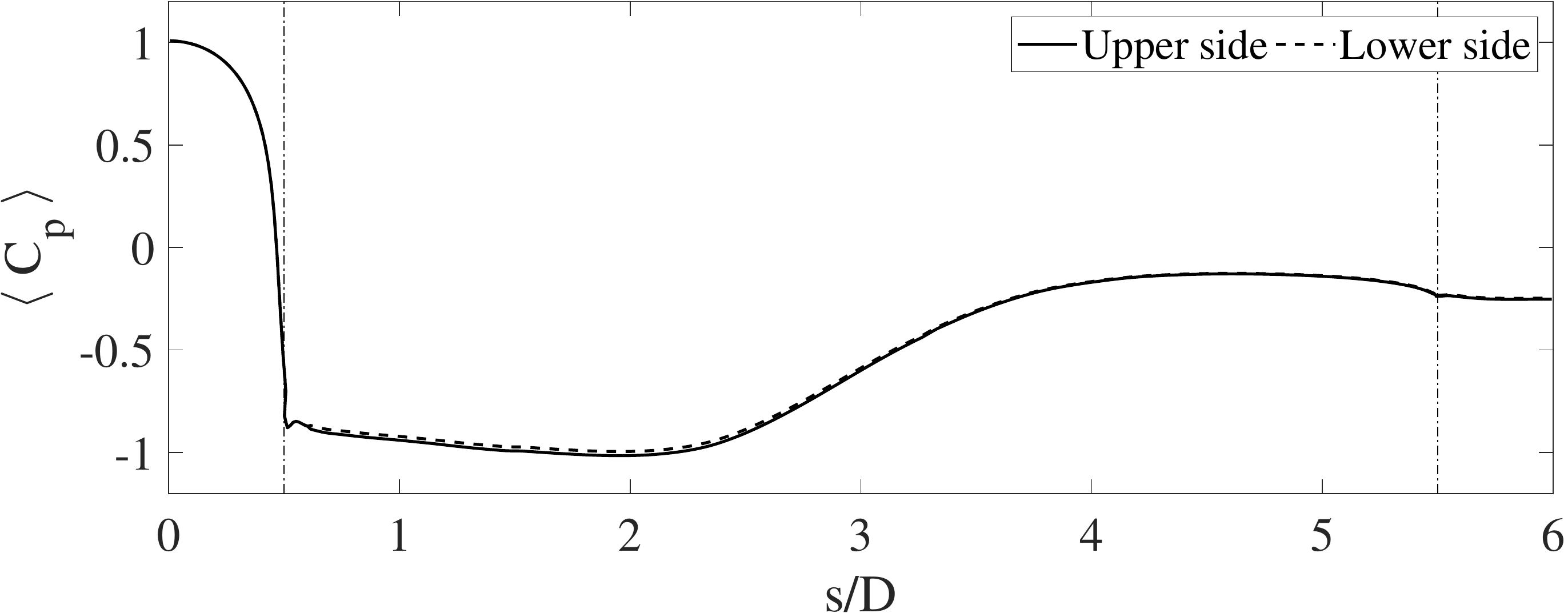}\\(a)\\
		\includegraphics[width=.65\textwidth]{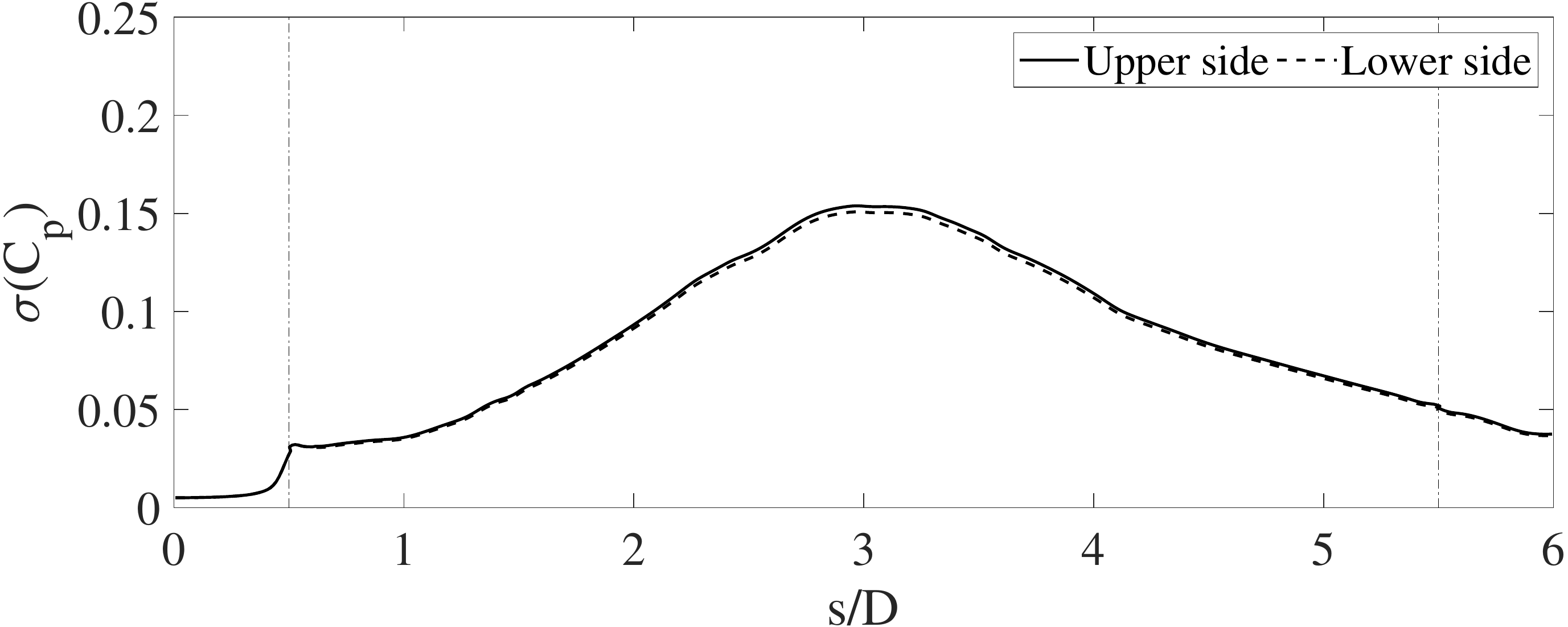}\\(b)\\
		\caption{(a) Pressure coefficient averaged in time and in the spanwise direction. (b) Standard deviation of the pressure coefficient, averaged in time and in the spanwise direction. Results for the upper and lower sides of the \textit{data-augmented LES}.}
		\label{fig:CP_distr_updown}
	\end{figure}

\begin{figure}
		\centering
		\includegraphics[width=.65\textwidth]{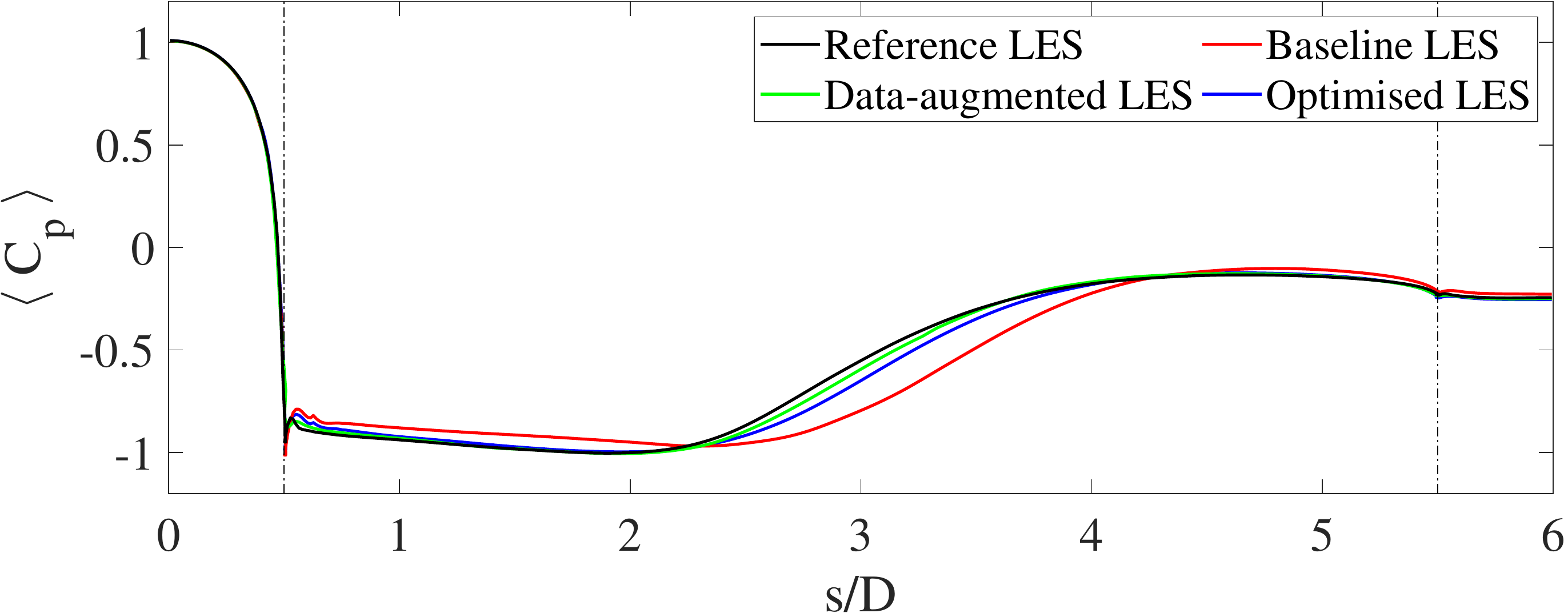}\\(a)\\
		\includegraphics[width=.65\textwidth]{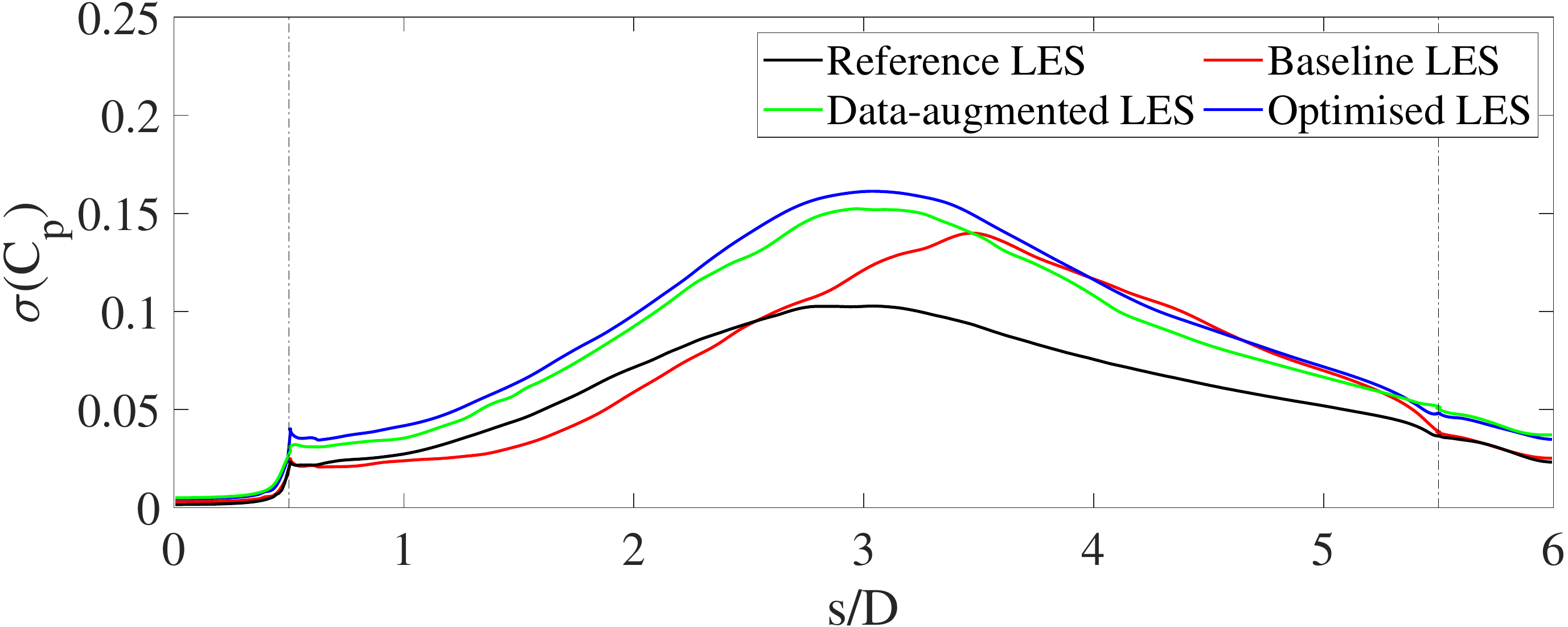}\\(b)\\
		\caption{(a) Pressure coefficient, $\langle C_p \rangle$, averaged in time, in the spanwise direction and between the upper and the lower surfaces of the cylinder. (b) Standard deviation of the pressure coefficient, $\sigma(C_p)$, averaged in space with the same criteria presented for $C_p$. The results for the \textit{data-augmented LES} and for the \textit{optimised LES} are compared with the ones for the \textit{baseline LES} and for the \textit{reference LES}.}
		\label{fig:CP_distr}
	\end{figure}

The excellent symmetry of the flow statistics is confirmed also by Figure \ref{fig:CP_distr_updown} showing the distribution of the mean value and of the standard deviation of $C_p$, obtained by averaging only in time and in the spanwise direction, along the upper and lower side of the cylinder. Improvement in the accuracy of the \textit{DA augmented LES} can be also observed in the results in Fig.~\ref{fig:CP_distr} (a),  where the mean pressure coefficient, $\langle C_p \rangle$, distributions obtained with the \textit{DA augmented LES} and the \textit{optimized LES} are compared with those of the \textit{baseline} and \textit{reference} simulations. The predicted profile for both the \textit{DA augmented LES} and the \textit{optimized LES} is significantly closer to the reference solution when compared with the \textit{baseline LES}. In addition, Fig.~\ref{fig:CP_distr} (b) also shows an improvement for the prediction of the variance of the pressure coefficient $\sigma(C_p)$. Despite differences in magnitude can be observed, the location of the peak, which is directly connected with the length of the mean recirculation, is well captured. This difficulty in capturing different orders of statistical moments via DA has already been observed in numerical simulation \citep{Mons2021_prf} and it is due to the features of the optimized model. In the present case, the optimization was performed over a global constant controlling an explicit filter, which controls the total dissipation. \cite{Moldovan2022} recently showed that both diffusive and dispersive errors must be controlled to provide a very accurate state estimation. Therefore, while the present results are satisfying considering the simplicity of the model used, one should also keep in mind that higher accuracy can be obtained with more sophisticated approaches. One can also see that results from the DA augmented simulation appear to be marginally better than the ones from the optimized simulation, on average.

\begin{figure}
		\centering
		\includegraphics[width=.65\textwidth]{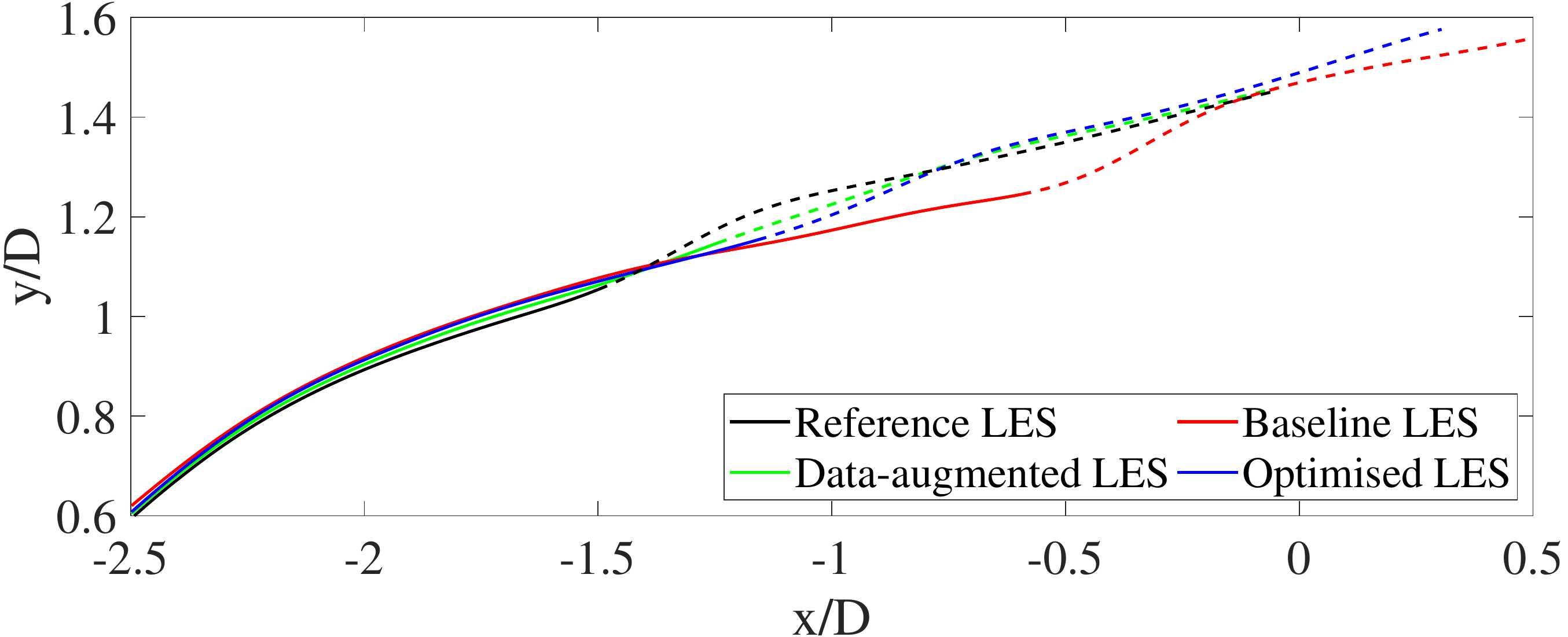}\\(a)\\
		\includegraphics[width=.65\textwidth]{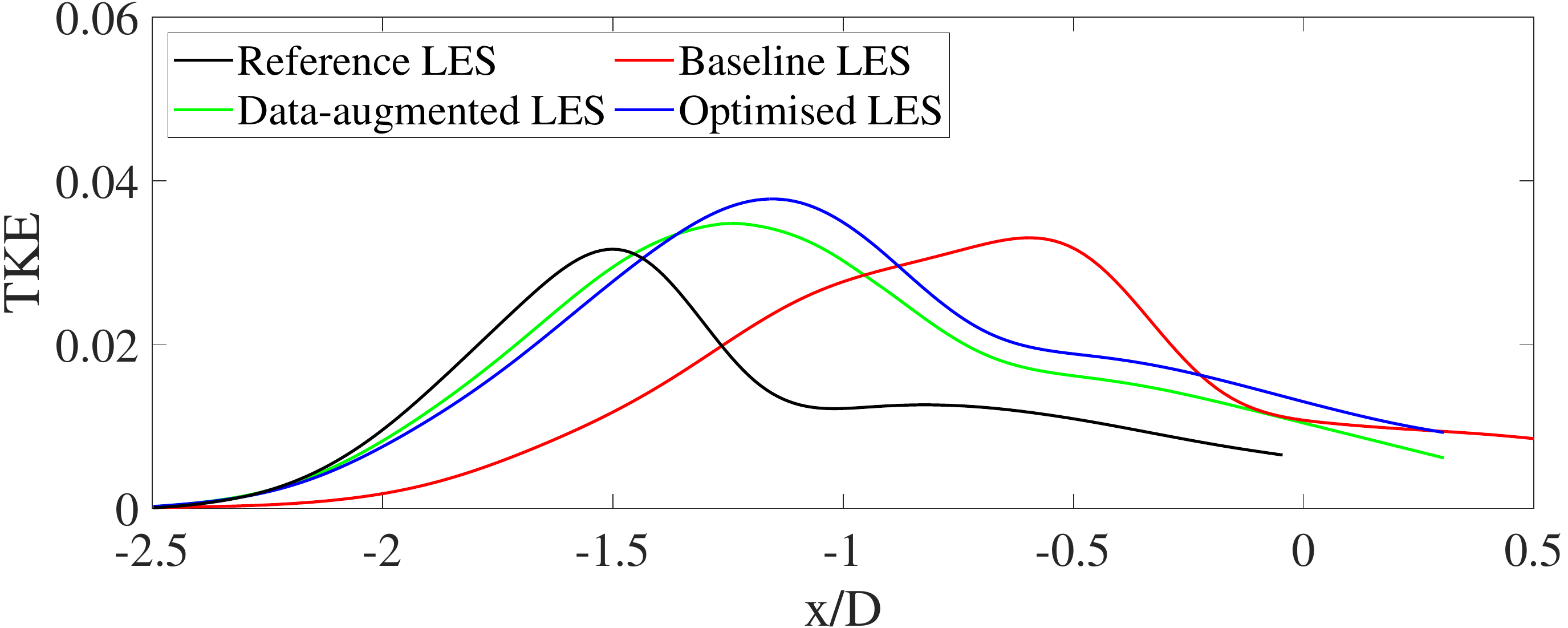}\\(b)\\
		\includegraphics[width=.65\textwidth]{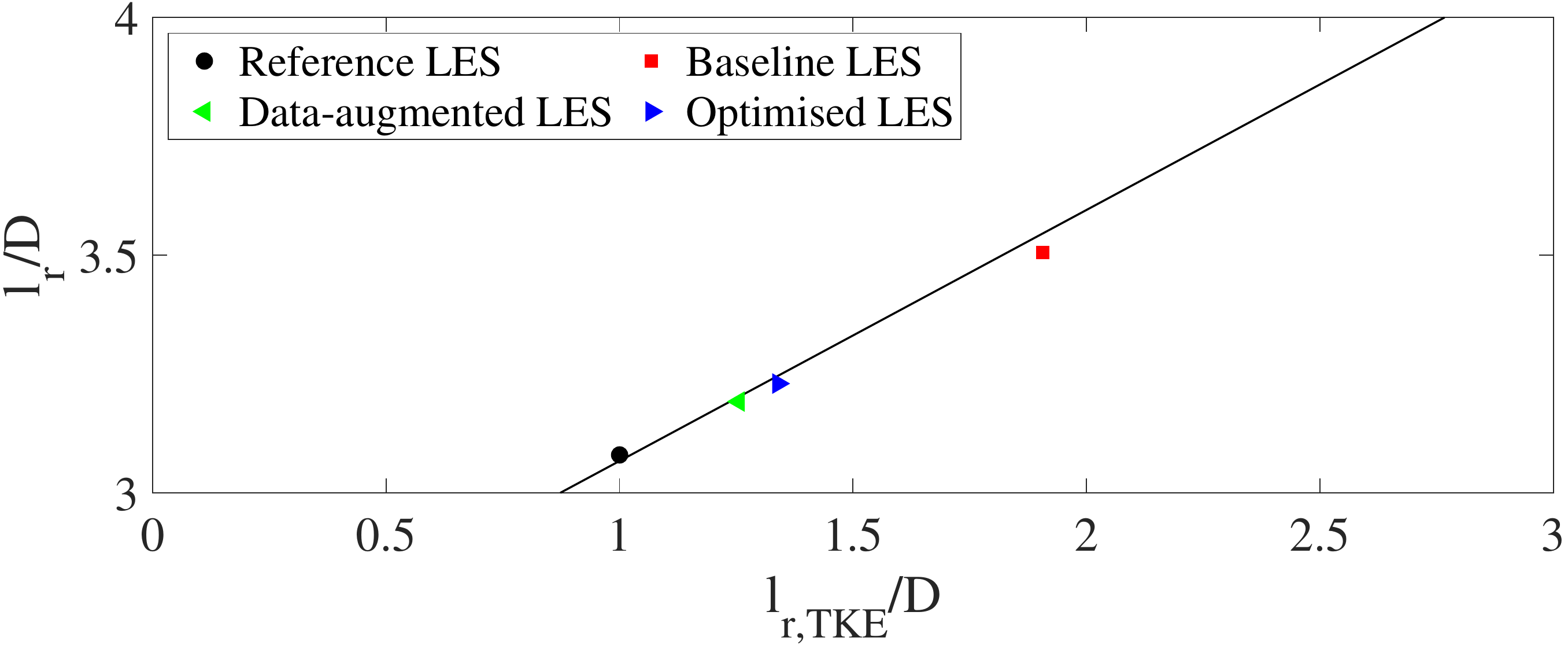}\\(c)\\
		\caption{(a) Position of the outer border of the detaching shear layer, (b) distribution of the $\textup{TKE}$ along the outer border of the detaching shear layer, and (c) correlation between the length of the mean recirculation region, $l_r$, and the position of maximum TKE on the shear layer border evaluated from the upstream corner, $l_{\textup{r,TKE}}$. The results for the \textit{data-augmented LES} and for the \textit{optimised LES} are compared with the ones for the \textit{baseline LES} and for the \textit{reference LES}.}
		\label{fig:DA_border}
	\end{figure}

The previous results show that both the \textit{DA augmented LES} and the \textit{optimized LES} give an overall accurate prediction of the mean flow topology on the cylinder side. As previously discussed, this is related with the growth of TKE along the detaching shear-layer. In Fig.~\ref{fig:DA_border}b, profiles for the turbulent kinetic energy  are shown along the shear-layer border, identified from the mean velocity field as in \cite{Rocchio2020} and shown in Fig.~\ref{fig:DA_border}a. Once again, one can see that the distribution of the TKE for the \textit{DA augmented LES} and the \textit{optimized LES} is qualitatively much more similar to the reference simulation when compared with the prior, consistently with the better prediction of the mean recirculation length. The correlation between the length of the mean recirculation region, $l_r$, and the position of maximum TKE on the shear layer border, $l_{r,TKE}$ is indeed shown in Fig.~\ref{fig:DA_border}c. The \textit{DA augmented} and the \textit{optimized} LES satisfy the correlation observed in \cite{Rocchio2020} and \cite{Lunghi2022}, and in better quantitative agreement with the \textit{reference LES}, as previously remarked.

\begin{figure}
		\centering
		\includegraphics[width=.45\textwidth]{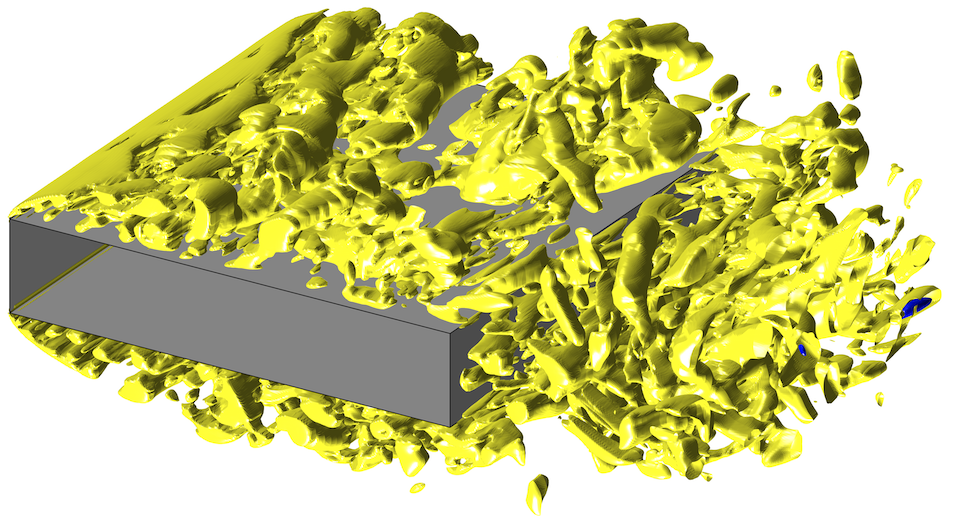}
		\includegraphics[width=.44\textwidth]{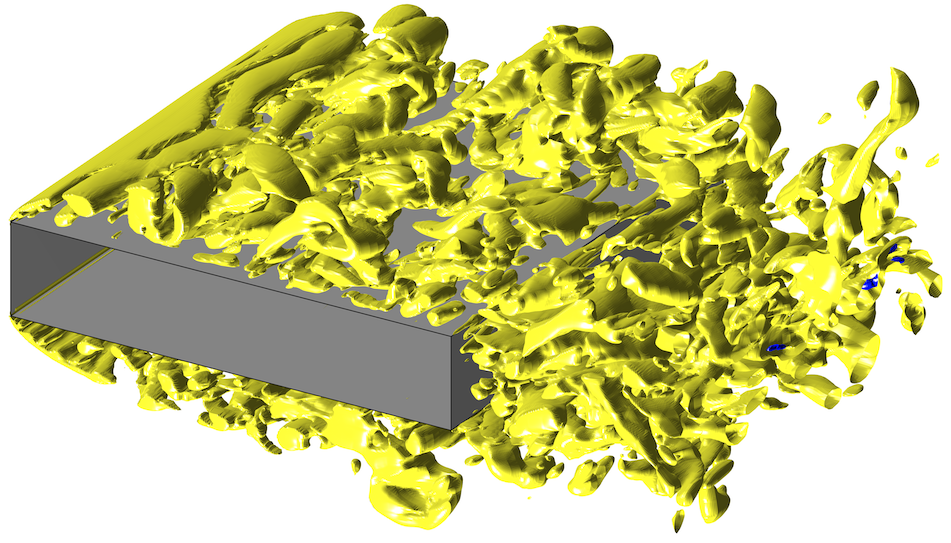}\\
		(a) \hspace{6cm} (b)\\
		\includegraphics[width=.49\textwidth]{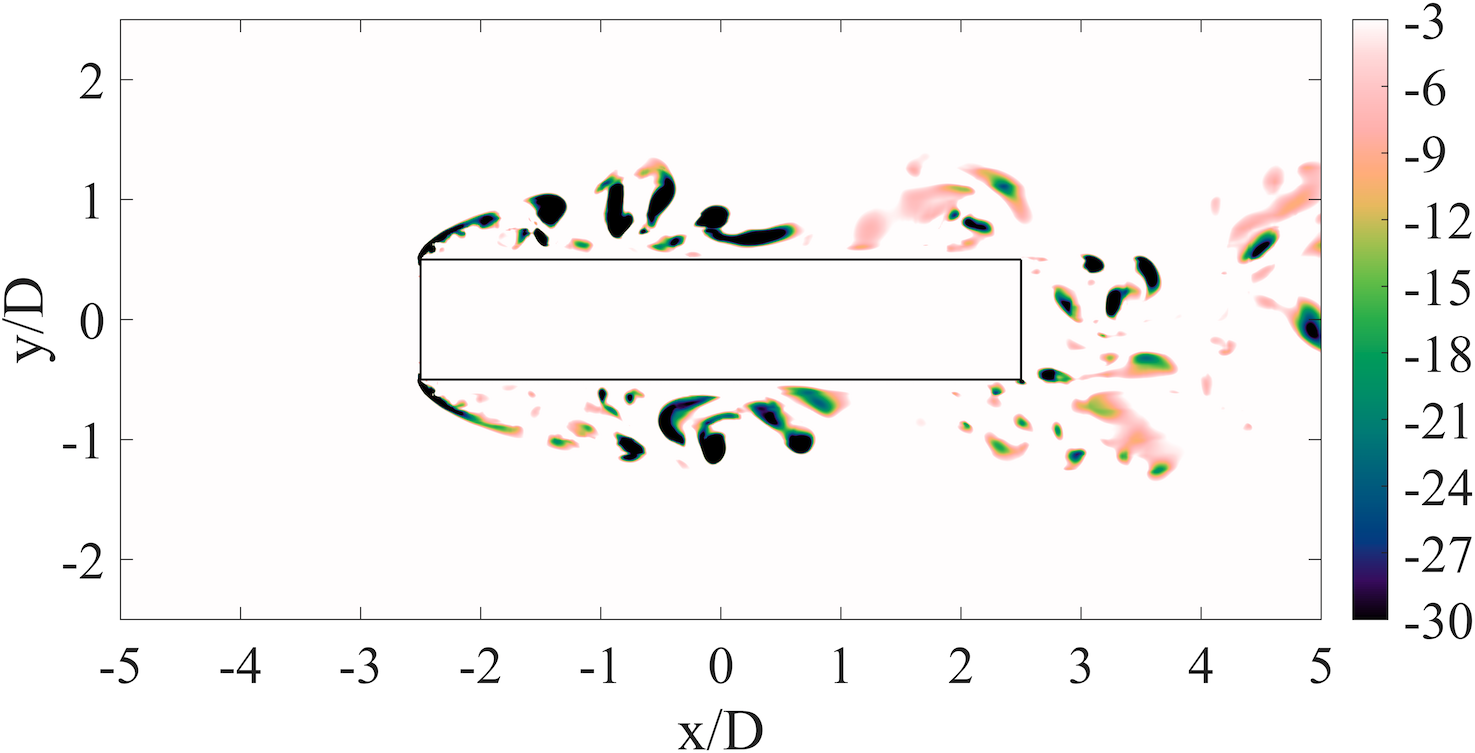}
		\includegraphics[width=.49\textwidth]{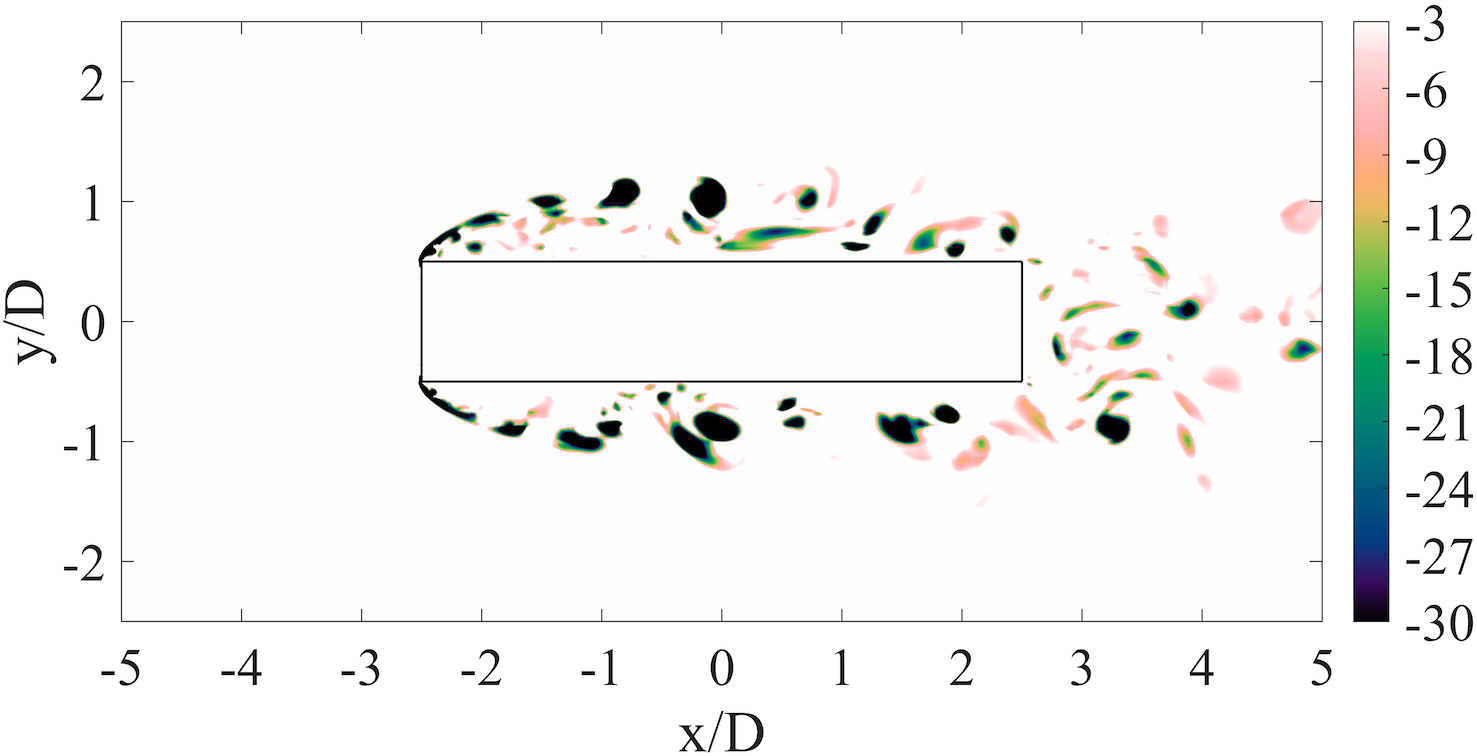}\\
		(c) \hspace{6cm} (d)\\
		\caption{(a,b) Isocontours of the instantaneous vortex indicator $\lambda_2$ and (c,d) its distribution on the spanwise symmetry plane ($z/D=0$). Results for (a,c) the \textit{data-augmented LES} and (b,d) the \textit{optimised LES}.}
		\label{fig:DA_l2}
	\end{figure}

Moreover, Figures \ref{fig:DA_l2}(a)-(d) show the instantaneous vortical structures for the \textit{DA augmented} and the \textit{optimized} LES. By comparing them with those of the \textit{reference} LES in Fig. \ref{fig:ref_l2}, a good agreement can be observed for the location of the instability and roll-up of the shear-layers.

A last remark should be provided about the optimized value of $w$ found in the DA procedure. It is significantly lower than the one used for the \textit{reference} LES. This is consistent with the effect of this parameter highlighted in \cite{Mariotti2017}: for a fixed grid resolution, reducing $w$ leads to a shorter mean recirculation. Since the \textit{baseline LES}, with the same $w$ as in the \textit{reference} one, gives a longer mean recirculation length, $w$ must be reduced in order to improve the agreement with the \textit{reference} simulation.

%
	
    \begin{table}
		\centering
		\begin{tabular}{cccc} \hline
			Bulk parameters & \textit{data-augmented LES} & \textit{optimised LES} \\ \hline
			$\langle C_D \rangle$ & 0.987 & 0.996 \\
			$\sigma(C_L)$       & 0.248 & 0.252 \\
			$St$                & 0.150 & 0.150 \\ \hline
		\end{tabular}
	\caption{Time-averaged horizontal-force coefficient, $\langle C_D \rangle$, standard deviation of the vertical-force coefficient, $\sigma(C_L)$, and Strouhal number, $St$, for the \textit{data-augmented LES} and for the \textit{optimised LES}.}
	\end{table}
	
\section{Conclusions}
\label{sec:conclusions}

The MGEnKF was tested and validated against the international benchmark referred to as BARC (Benchmark of the Aerodynamics of a Rectangular 5:1 Cylinder). 
Despite the simple geometry, the flow shows complex features characterized by shear-layer separation from the upstream edges, unsteady reattachment on the cylinder side and vortex shedding in the wake. Thus, BARC is a paradigmatic example of the intricate flows around elongated bodies often encountered in civil engineering applications. 


A Data Assimilation experiment was performed where the MGEnKF algorithm is used to improve the predictive capabilities of LES (model), by integrating sparse high-fidelity data (observation) from a LES carried out on  a very refined grid, which is considered here as the ground truth. The MGEnKF procedure is based on a cluster of simulations, namely i) one main simulation performed using the same mesh used for the prior (baseline grid LES) and ii) an ensemble of simulations, which are responsible for the state estimation and the parametric optimization, run on a coarse grid. The DA strategy aims for the optimization of a global parameter, $w$, that regulates the numerical dissipation introduced by a modal filter. As in \cite{Mariotti2017, Rocchio2020}, this dissipation may also be considered as a SGS dissipation. The working hypothesis was that the additional numerical errors that appear in LES on coarser grids can be reduced by calibrating the additional introduced dissipation. Besides the optimization step, the statistics obtained from the coarse grid ensemble are used to update the main LES with a KF correction step.

The performance of the MGEnKF estimator has been assessed by evaluating first and second order statistics of velocity and pressure. More specifically, the results of four different LES simulations have been compared: the \textit{reference LES}, the \textit{baseline LES}, run on a coarser grid, the \textit{data-augmented LES} and the \textit{optimized LES}, carried out with the optimal value of $w$ given by the MGEnKF procedure. Both the \textit{data-augmented} and the \textit{optimized} LES are run on the same grid as the \textit{baseline} one. The results of the \textit{baseline LES} show significant discrepancies compared to the \textit{reference one}. Conversely, the predictions obtained for both the \textit{data-augmented} and the \textit{optimized} LES are in very good agreement for all the physical quantities investigated.

One striking observation is that, despite the lack of symmetry in the distribution of the sensors providing observation and a rather small observation region, the data-driven tool was able to provide perfectly symmetric statistics and a global accurate representation of the flow features even far from the sensors. Although only velocity signals were provided by the sensors, the procedure was able to provide an accurate prediction of the pressure statistics. These results indicate that, despite the larger computational costs, the use of complex models such as CFD codes may be as attractive as very fast reduced order models for data-driven applications and for machine-learning.  Indeed, a complex model able to independently provide physical information may be more suitable for integration of highly sparse data in space and time, opening the possibility of using sparse experimental data as sensor information. Moreover, complex models exhibit a lower sensitivity to problems of convergence or overfitting, which are quite regularly observed for very simplified models.

In future works, the sequential DA procedure used in this work can be substantially improved using the full MGEnKF algorithm, exploiting the beneficial effects of the \textit{inner loop} which have been recently investigated in the work by \cite{Moldovan2022}. The \textit{inner loop} \textit{trains} a \textit{model correction term} integrated in the simulations run on the coarse grid level, using surrogate observation extracted from the higher resolution simulation run within the MGEnKF. The objective of the \textit{model correction term} is to reduce the discrepancy between fine and coarse grid levels, in order to provide a smoother transition of the DA results between the multilevel representation. The main challenge here is deriving a consistent form for the \textit{model correction term}, potentially using machine learning algorithms.

This work was performed in the framework of the HPCEuropa3 project HPC17ICRK0. Also, computational resources of the project EDARI A0012A07590 have been used to perform calculations. Benedetto Rochio is warmly acknowledged for the help he provided for the set up of the numerical simulations. The \textit{Direction Générale de l'Armement} (2018-0054 REPUB DGA) and the \textit{Région Nouvelle Aquitaine} (2018-0042 NAQ) are warmly acknowledged for their funding.

The authors report no conflict of interest

\clearpage
\newpage

\bibliographystyle{jfm}
\bibliography{Bibliography_HDR}

\begin{thebibliography}{52}
\expandafter\ifx\csname natexlab\endcsname\relax\def\natexlab#1{#1}\fi
\def\au#1{#1} \def\ed#1{#1} \def\yr#1{#1}\def\at#1{#1}\def\jt#1{\textit{#1}}
  \def\bt#1{#1}\def\bvol#1{\textbf{#1}} \def\vol#1{#1} \def\pg#1{#1}
  \def\publ#1{#1}\def\arxiv#1{#1}\def\org#1{#1}\def\st#1{\textit{#1}}

\bibitem[Asch {\em et~al.\/}(2016)Asch, Bocquet \& Nodet]{Asch2016_SIAM}
{\sc \au{Asch, M.}, \au{Bocquet, M.} \& \au{Nodet, M.}} \yr{2016} {\em {Data
  Assimilation: methods, algorithms, and applications}\/}.  \publ{SIAM}.

\bibitem[Brandt(1977)]{Brandt1977_mc}
{\sc \au{Brandt, A.}} \yr{1977}  \at{{Multi-level adaptive solutions to
  boundary-value problems}}.  \jt{Mathematics of Computation}  \bvol{31},
  \pg{333--390}.

\bibitem[Bruno {\em et~al.\/}(2012)Bruno, Coste \& Fransos]{Bruno2012}
{\sc \au{Bruno, L.}, \au{Coste, N.} \& \au{Fransos, D.}} \yr{2012}
  \at{Simulated flow around a rectangular 5:1 cylinder: Spanwise discretisation
  effects and emerging flow features}.  \jt{J. Wind Eng. Ind. Aerodyn.}
  \bvol{104-106},  \pg{203--215}.

\bibitem[Bruno {\em et~al.\/}(2014)Bruno, Salvetti \& Ricciardelli]{Bruno2014}
{\sc \au{Bruno, L.}, \au{Salvetti, M.~V.} \& \au{Ricciardelli, F.}} \yr{2014}
  \at{{B}enchmark on the {A}erodynamics of a {R}ectangular 5:1 {C}ylinder: an
  overview after the first four years of activity}.  \jt{J. Wind Eng. Ind.
  Aerodyn.}  \bvol{126},  \pg{87--106}.

\bibitem[Burgers {\em et~al.\/}(1998)Burgers, Van~Leeuwen \&
  Evensen]{Burgers1998}
{\sc \au{Burgers, Gerrit}, \au{Van~Leeuwen, Peter~Jan} \& \au{Evensen, Geir}}
  \yr{1998}  \at{On the analysis scheme in the ensemble kalman filter}.
  \jt{Monthly Weather Review}  \bvol{126}.

\bibitem[Chandramouli {\em et~al.\/}(2020)Chandramouli, Memin \&
  Heitz]{Chandramouli_Memin_Heitz_JCP_2020}
{\sc \au{Chandramouli, P.}, \au{Memin, E.} \& \au{Heitz, D.}} \yr{2020}  \at{4d
  large scale variational data assimilation of a turbulent flow with a dynamics
  error model}.  \jt{J. Comp. Phys.}  \bvol{412},  \pg{109446}.

\bibitem[Chernov {\em et~al.\/}(2021)Chernov, Hoel, Law, Nobile \&
  Tempone]{KodyLaw2020}
{\sc \au{Chernov, Alexey}, \au{Hoel, Håkon}, \au{Law, Kody J.~H.}, \au{Nobile,
  Fabio} \& \au{Tempone, Raul}} \yr{2021}  \at{{Multilevel ensemble Kalman
  filtering for spatio-temporal processes}}.  \jt{Numerische Mathematik}
  \bvol{147},  \pg{71--125}.

\bibitem[Daley(1991)]{Daley1991_cambridge}
{\sc \au{Daley, S.~B.}} \yr{1991} {\em {Atmospheric Data Analysis}\/}.
  \publ{Cambridge University Press}.

\bibitem[Domarazdki(2010)]{nek_sgs_1}
{\sc \au{Domarazdki, J.~A.}} \yr{2010}  \at{Large eddy simulations without
  explicit eddy viscosity models}.  \jt{Int. J. Comput. Fluid Dyn.}
  \bvol{24(10)},  \pg{435--447}.

\bibitem[Duchaine {\em et~al.\/}(2015)Duchaine, Jaur{\'{e}}, Poitou,
  Qu{\'{e}}merais, Staffelbach, Morel \& Gicquel]{Duchaine_2015}
{\sc \au{Duchaine, Florent}, \au{Jaur{\'{e}}, St{\'{e}}phan}, \au{Poitou,
  Damien}, \au{Qu{\'{e}}merais, Eric}, \au{Staffelbach, Gabriel}, \au{Morel,
  Thierry} \& \au{Gicquel, Laurent}} \yr{2015}  \at{Analysis of high
  performance conjugate heat transfer with the {OpenPALM} coupler}.
  \jt{Computational Science \& Discovery}  \bvol{8}~(1),  \pg{015003}.

\bibitem[Evensen(1994)]{Evensen1994}
{\sc \au{Evensen, Geir}} \yr{1994}  \at{Sequential data assimilation with a
  nonlinear quasi-geostrophic model using monte-carlo methods to forecast error
  statistics}.  \jt{Journal of Geophysical Research}  \bvol{99},
  \pg{10143--10162}.

\bibitem[Evensen(2009)]{Evensen2009_Springer}
{\sc \au{Evensen, G.}} \yr{2009} {\em {Data Assimilation: The Ensemble Kalman
  Filter}\/}.  \publ{Springer-Verlag/Berlin/Heildelberg}.

\bibitem[Fischer {\em et~al.\/}(2008)Fischer, Lottes \&
  Kerkemeier]{fischer2008}
{\sc \au{Fischer, P.~F.}, \au{Lottes, J.~W.} \& \au{Kerkemeier, S.G.}}
  \yr{2008}  \at{{nek5000 Web page}}.  \jt{{http://nek5000.mcs.anl.gov}} .

\bibitem[Fossum {\em et~al.\/}(2020)Fossum, Mannseth \& Stordal]{Fossum2020_cg}
{\sc \au{Fossum, K.}, \au{Mannseth, T.} \& \au{Stordal, A.~S.}} \yr{2020}
  \at{Assessment of multilevel ensemble-based data assimilation for reservoir
  history matching}.  \jt{Computational Geosciences}  \bvol{24},
  \pg{217--239}.

\bibitem[Gaspari \& Cohn(1999)]{Gaspari_1999}
{\sc \au{Gaspari, G.} \& \au{Cohn, S.~E.}} \yr{1999}  \at{Construction of
  correlation functions in two and three dimensions}.  \jt{Quarterly Journal of
  the Royal Meteorological Society}  \bvol{125}~(554),  \pg{723--757},
  \arxiv{arXiv:
  https://rmets.onlinelibrary.wiley.com/doi/pdf/10.1002/qj.49712555417}.

\bibitem[Geurts(2009)]{Geurts2009}
{\sc \au{Geurts, Bernard~J.}} \yr{2009}  \at{Analysis of errors occurring in
  large eddy simulation}.  \jt{Philosophical Transactions of the Royal Society
  A: Mathematical, Physical and Engineering Sciences}  \bvol{367},  \pg{2873 --
  2883}.

\bibitem[Gorodetsky {\em et~al.\/}(2020)Gorodetsky, Geraci, Eldred \&
  Jakeman]{Gorodetsky2020_jcp}
{\sc \au{Gorodetsky, A.~A.}, \au{Geraci, G.}, \au{Eldred, M.~S.} \&
  \au{Jakeman, J.~D.}} \yr{2020}  \at{A generalized approximate control variate
  framework for multifidelity uncertainty quantification}.  \jt{Journal of
  Computational Physics}  \bvol{408},  \pg{109257}.

\bibitem[Hoel {\em et~al.\/}(2016)Hoel, Law \& Tempone]{Hoel2016_SIAM}
{\sc \au{Hoel, H.}, \au{Law, K. J.~H.} \& \au{Tempone, R.}} \yr{2016}
  \at{{Multilevel ensemble Kalman filtering}}.  \jt{SIAM J. Numer. Anal.}
  \bvol{54(3)},  \pg{1813--1839}.

\bibitem[Kalman(1960)]{Kalman1960_jbe}
{\sc \au{Kalman, R.~E.}} \yr{1960}  \at{{A new approach to linear filtering and
  prediction problems}}.  \jt{Journal of Basic Engineering}  \bvol{82},
  \pg{35--45}.

\bibitem[Labahn {\em et~al.\/}(2019)Labahn, Wu, Coriton, Frank \&
  Ihme]{Labahn2019_pci}
{\sc \au{Labahn, J.~W.}, \au{Wu, H.}, \au{Coriton, B.}, \au{Frank, J.~H.} \&
  \au{Ihme, M.}} \yr{2019}  \at{{Data assimilation using high-speed
  measurements and LES to examine local extinction events in turbulent
  flames}}.  \jt{Proceedings of the Combustion Institute}  \bvol{37},
  \pg{2259--2266}.

\bibitem[Lucor {\em et~al.\/}(2007)Lucor, Meyers \& Sagaut]{Lucor2007_jfm}
{\sc \au{Lucor, D.}, \au{Meyers, J.} \& \au{Sagaut, P.}} \yr{2007}
  \at{{Sensitivity analysis of large-eddy simulations to subgrid-scale-model
  parametric uncertainty using polynomial chaos}}.  \jt{Journal of Fluid
  Mechanics}  \bvol{585},  \pg{255--279}.

\bibitem[Lunghi {\em et~al.\/}(2022)Lunghi, Pasqualetto, Rocchio, Mariotti \&
  Salvetti]{Lunghi2022}
{\sc \au{Lunghi, G.}, \au{Pasqualetto, E.}, \au{Rocchio, B.}, \au{Mariotti, A.}
  \& \au{Salvetti, M.~V.}} \yr{2022}  \at{{Impact of the lateral mean
  recirculation characteristics on the near wake and bulk quantities of the
  BARC configuration}}.  \jt{Wind and Structures}  \bvol{34(1)},
  \pg{115--125}.

\bibitem[Maday {\em et~al.\/}(1990)Maday, Patera \& R{\o}nquist]{mady1990}
{\sc \au{Maday, Y.}, \au{Patera, A.~T.} \& \au{R{\o}nquist, E.~M.}} \yr{1990}
  \at{An operator-integration-factor splitting method for time-dependent
  problems: Application to incompressible fluid flow.}  \jt{J. Sci. Comput.}
  \bvol{5 (4)},  \pg{263--292}.

\bibitem[Mariotti {\em et~al.\/}(2017)Mariotti, Siconolfi \&
  Salvetti]{Mariotti2017}
{\sc \au{Mariotti, A.}, \au{Siconolfi, L.} \& \au{Salvetti, M.~V.}} \yr{2017}
  \at{Stochastic sensitivity analysis of large-eddy simulation predictions of
  the flow around a 5:1 rectangular cylinder}.  \jt{Eur. J. Mech. B-Fluids}
  \bvol{62},  \pg{149--165}.

\bibitem[Mathew {\em et~al.\/}(2003)Mathew, Lechner, Foysi, Setsterhenn \&
  Friedrich]{nek_sgs_2}
{\sc \au{Mathew, J.}, \au{Lechner, R.}, \au{Foysi, H.}, \au{Setsterhenn, J.} \&
  \au{Friedrich, R.}} \yr{2003}  \at{An explicit filtering method for large
  eddy simulation of compressible flows}.  \jt{Phys. Fluids}  \bvol{15},
  \pg{2279--2289}.

\bibitem[Meldi {\em et~al.\/}(2011)Meldi, Lucor \& Sagaut]{Meldi2011_pof}
{\sc \au{Meldi, M.}, \au{Lucor, D.} \& \au{Sagaut, P.}} \yr{2011}  \at{{Is the
  Smagorinsky coefficient sensitive to uncertainty in the form of the energy
  spectrum?}}  \jt{Physics of Fluids}  \bvol{23},  \pg{125109}.

\bibitem[Meldi \& Poux(2017)]{Meldi2017_jcp}
{\sc \au{Meldi, M.} \& \au{Poux, A.}} \yr{2017}  \at{{A reduced order model
  based on Kalman Filtering for sequential Data Assimilation of turbulent
  flows}}.  \jt{Journal of Computational Physics}  \bvol{347},  \pg{207--234}.

\bibitem[Meldi {\em et~al.\/}(2012)Meldi, Salvetti \& Sagaut]{Meldi2012_pof}
{\sc \au{Meldi, M.}, \au{Salvetti, M.~V.} \& \au{Sagaut, P.}} \yr{2012}
  \at{{Quantification of errors in large-eddy simulations of a spatially
  evolving mixing layer using polynomial chaos}}.  \jt{Physics of Fluids}
  \bvol{24},  \pg{035101}.

\bibitem[Meyers {\em et~al.\/}(2003)Meyers, Geurts \& Baelmans]{Meyers2003}
{\sc \au{Meyers, Johan}, \au{Geurts, Bernard~J.} \& \au{Baelmans, Martine}}
  \yr{2003}  \at{Database analysis of errors in large-eddy simulation}.
  \jt{Physics of Fluids}  \bvol{15}~(9),  \pg{2740--2755},  \arxiv{arXiv:
  https://doi.org/10.1063/1.1597683}.

\bibitem[Meyers \& Sagaut(2006)]{Meyers2006_jfm}
{\sc \au{Meyers, J.} \& \au{Sagaut, P.}} \yr{2006}  \at{{On the model
  coefficients for the standard and the variational multi-scale Smagorinsky
  model}}.  \jt{Journal of Fluid Mechanics}  \bvol{569},  \pg{287--319}.

\bibitem[Moldovan {\em et~al.\/}(2021)Moldovan, Lehnasch, Cordier \&
  Meldi]{Moldovan2021_jcp}
{\sc \au{Moldovan, G.}, \au{Lehnasch, G.}, \au{Cordier, L.} \& \au{Meldi, M.}}
  \yr{2021}  \at{{A multigrid/ensemble Kalman Filter strategy for assimilation
  of unsteady flows}}.  \jt{Journal of Computational Physics}  \bvol{443},
  \pg{110481}.

\bibitem[Moldovan {\em et~al.\/}(2022)Moldovan, Lehnasch, Cordier \&
  Meldi]{Moldovan2022}
{\sc \au{Moldovan, G.}, \au{Lehnasch, G.}, \au{Cordier, L.} \& \au{Meldi, M.}}
  \yr{2022}  \at{Optimized parametric inference for the inner loop of the
  multigrid ensemble kalman filter}.  \jt{Journal of Computational Physics}
  \pg{p. 111621}.

\bibitem[Mons {\em et~al.\/}(2021)Mons, Du \& Zaki]{Mons2021_prf}
{\sc \au{Mons, V.}, \au{Du, Y.} \& \au{Zaki, T.}} \yr{2021}
  \at{Ensemble-variational assimilation of statistical data in large-eddy
  simulation}.  \jt{Physical Review Fluids}  \bvol{6},  \pg{104607}.

\bibitem[Mons {\em et~al.\/}(2019)Mons, Wang \& Zaki]{Mons2019_jcp}
{\sc \au{Mons, V.}, \au{Wang, Q.} \& \au{Zaki, T.~A.}} \yr{2019}
  \at{Kriging-enhanced ensemble variational data assimilation for scalar-source
  identification in turbulent environments}.  \jt{Journal of Computational
  Physics}  \bvol{398},  \pg{108856}.

\bibitem[Moradkhani {\em et~al.\/}(2005)Moradkhani, Sorooshian, Gupta \&
  Houser]{moradkhani2005}
{\sc \au{Moradkhani, Hamid}, \au{Sorooshian, Soroosh}, \au{Gupta, Hoshin~V.} \&
  \au{Houser, Paul~R.}} \yr{2005}  \at{Dual state-parameter estimation of
  hydrological models using ensemble {Kalman} filter}.  \jt{Advances in Water
  Resources}  \bvol{28}~(2),  \pg{135--147}.

\bibitem[Ouvrard {\em et~al.\/}(2010)Ouvrard, Koobus, Dervieux \&
  Salvetti]{Ouvrard2010}
{\sc \au{Ouvrard, Hilde}, \au{Koobus, Bruno}, \au{Dervieux, Alain} \&
  \au{Salvetti, Maria~Vittoria}} \yr{2010}  \at{Classical and variational
  multiscale les of the flow around a circular cylinder on unstructured grids}.
   \jt{Computers \& Fluids}  \bvol{39}~(7),  \pg{1083--1094}.

\bibitem[Pope(2000)]{Pope2000_cambridge}
{\sc \au{Pope, S.~B.}} \yr{2000} {\em Turbulent flows\/}.  \publ{Cambridge
  University Press}.

\bibitem[Popov {\em et~al.\/}(2021)Popov, Mou, Sandu \&
  Iliescu]{Popov2021_SIAM}
{\sc \au{Popov, A.~A.}, \au{Mou, C.}, \au{Sandu, A.} \& \au{Iliescu, T.}}
  \yr{2021}  \at{{A Multifidelity Ensemble Kalman Filter with Reduced Order
  Control Variates}}.  \jt{SIAM Journal on Scientific Computing}  \bvol{43},
  \pg{A1134--A1162}.

\bibitem[Rezaeiravesh \& Liefvendahl(2018)]{Rezaeiravesh2018}
{\sc \au{Rezaeiravesh, S.} \& \au{Liefvendahl, M.}} \yr{2018}  \at{Effect of
  grid resolution on large eddy simulation of wall-bounded turbulence}.
  \jt{Physics of Fluids}  \bvol{30}~(5).

\bibitem[Rezaeiravesh {\em et~al.\/}(2021)Rezaeiravesh, Vinuesa \&
  Schlatter]{Rezaeiravesh2021}
{\sc \au{Rezaeiravesh, S.}, \au{Vinuesa, R.} \& \au{Schlatter, P.}} \yr{2021}
  \at{On numerical uncertainties in scale-resolving simulations of canonical
  wall turbulence}.  \jt{Computers and Fluids}  \bvol{227}.

\bibitem[Rocchio {\em et~al.\/}(2020)Rocchio, Mariotti \&
  Salvetti]{Rocchio2020}
{\sc \au{Rocchio, B.}, \au{Mariotti, A.} \& \au{Salvetti, M.~V.}} \yr{2020}
  \at{Flow around a 5:1 rectangular cylinder: Effects of upstream-edge
  rounding}.  \jt{J. Wind Eng. Ind. Aerodyn.}  \bvol{204},  \pg{104237}.

\bibitem[Rochoux {\em et~al.\/}(2015)Rochoux, Ricci, Lucor, Cuenot \&
  Trouve]{Rochoux2014_nhess}
{\sc \au{Rochoux, M.~C.}, \au{Ricci, S.}, \au{Lucor, D.}, \au{Cuenot, B.} \&
  \au{Trouve, A.}} \yr{2015}  \at{{Towards predictive data-driven simulations
  of wildfire spread - Part I: Reduced-cost Ensemble Kalman Filter based on a
  Polynomial Chaos surrogate model for parameter estimation}}.  \jt{Natural
  Hazards and Earth System Sciences}  \bvol{14},  \pg{2951--2973}.

\bibitem[Safta {\em et~al.\/}(2017)Safta, Blaylock, Templeton, Domino, Sargsyan
  \& Najm]{Safta2017376}
{\sc \au{Safta, C.}, \au{Blaylock, M.}, \au{Templeton, J.}, \au{Domino, S.},
  \au{Sargsyan, K.} \& \au{Najm, H.}} \yr{2017}  \at{{Uncertainty
  quantification in LES of channel flow}}.  \jt{International Journal for
  Numerical Methods in Fluids}  \bvol{83}~(4),  \pg{376--401}.

\bibitem[Sagaut(2005)]{Sagaut2006_springer}
{\sc \au{Sagaut, P.}} \yr{2005} {\em {Large-eddy simulation for incompressible
  flows. An introduction, third edition}\/}.  \publ{Springer-Verlag}.

\bibitem[Siripatana {\em et~al.\/}(2019)Siripatana, Giraldi, Le~Maître, Knio
  \& Hoteit]{Siripatana2019_cg}
{\sc \au{Siripatana, A.}, \au{Giraldi, L.}, \au{Le~Maître, O.~P.}, \au{Knio,
  O.~M.} \& \au{Hoteit, I.}} \yr{2019}  \at{{Combining ensemble Kalman filter
  and multiresolution analysis for efficient assimilation into adaptive mesh
  models}}.  \jt{Computational Geosciences}  \bvol{23},  \pg{1259--1276}.

\bibitem[Soares {\em et~al.\/}(2018)Soares, Maschio \& Schiozer]{SOARES2018110}
{\sc \au{Soares, Ricardo~Vasconcellos}, \au{Maschio, Célio} \& \au{Schiozer,
  Denis~José}} \yr{2018}  \at{{Applying a localization technique to Kalman
  Gain and assessing the influence on the variability of models in history
  matching}}.  \jt{Journal of Petroleum Science and Engineering}  \bvol{169},
  \pg{110--125}.

\bibitem[Tandeo {\em et~al.\/}(2020)Tandeo, Ailliot, Bocquet, Carrassi,
  Miyoshi, Pulido \&
  Zhen]{Tandeo_Ailliot_Bocquet_Carrassi_Miyoshi_Pulido_Zhen_MWR_2020}
{\sc \au{Tandeo, Pierre}, \au{Ailliot, Pierre}, \au{Bocquet, Marc},
  \au{Carrassi, Alberto}, \au{Miyoshi, Takemasa}, \au{Pulido, Manuel} \&
  \au{Zhen, Yicun}} \yr{2020}  \at{{Joint Estimation of Model and Observation
  Error Covariance Matrices in Data Assimilation: a Review}}.  \jt{Monthly
  Weather Review}  \bvol{148},  \pg{3973–3994}.

\bibitem[Wang \& Zaki(2022)]{wang_zaki_2022}
{\sc \au{Wang, Mengze} \& \au{Zaki, Tamer~A.}} \yr{2022}  \at{Synchronization
  of turbulence in channel flow}.  \jt{Journal of Fluid Mechanics}  \bvol{943}.

\bibitem[Wesseling \& Oosterlee(2001)]{Wesseling1999_jcam}
{\sc \au{Wesseling, P.} \& \au{Oosterlee, C.~W.}} \yr{2001}  \at{{Geometric
  multigrid with applications to computational fluid dynamics}}.  \jt{Journal
  of Computational and Applied Mathematics}  \bvol{128},  \pg{311--334}.

\bibitem[Xiao {\em et~al.\/}(2016)Xiao, Wu, Wang, Sun \& Roy]{Xiao2016_jcp}
{\sc \au{Xiao, H.}, \au{Wu, J.~L.}, \au{Wang, J.~X.}, \au{Sun, R.} \& \au{Roy,
  C.}} \yr{2016}  \at{{Quantifying and reducing model-form uncertainties in
  Reynolds-averaged Navier--Stokes simulations: A data-driven, physics informed
  Bayesian approach}}.  \jt{Journal of Computational Physics}  \bvol{324},
  \pg{115--136}.

\bibitem[Zhang {\em et~al.\/}(2020)Zhang, Xiao, Gomez \&
  Coutier-Delgosha]{Zhang2020_cf}
{\sc \au{Zhang, X.}, \au{Xiao, H.}, \au{Gomez, T.} \& \au{Coutier-Delgosha,
  O.}} \yr{2020}  \at{{Evaluation of ensemble methods for quantifying
  uncertainties in steady-state CFD applications with small ensemble sizes}}.
  \jt{Computers \& Fluids}  \bvol{203},  \pg{104530}.

\bibitem[Zhang {\em et~al.\/}(2021)Zhang, Xiao, He \& Wang]{Zhang2021_cf}
{\sc \au{Zhang, X.~L.}, \au{Xiao, H.}, \au{He, G.~W.} \& \au{Wang, S.~Z.}}
  \yr{2021}  \at{Assimilation of disparate data for enhanced reconstruction of
  turbulent mean flows}.  \jt{Computers \& Fluids}  \bvol{224},  \pg{104962}.

\end{thebibliography}

\end{document}